\documentclass[graybox, nosecnum]{svmult}

\usepackage{mathptmx}       
\usepackage{helvet}         
\usepackage{courier}        
\usepackage{type1cm}        
\usepackage{makeidx}         
\usepackage{graphicx}        
\usepackage{multicol}        
\usepackage[bottom]{footmisc}
\usepackage{hyperref}        
\usepackage{soul}            
\newcommand{\lambdabar}{{\mkern0.75mu\mathchar '26\mkern -9.75mu\lambda}}

\makeindex             

\begin{document}
\title*{Direct Nuclear Reactions}
\author{C.A. Bertulani \thanks{corresponding author} and A. Bonaccorso}
\institute{C.A. Bertulani \at Department of Physics and Astronomy, Texas A\&M University-Commerce, Commerce, TX 75429, USA, \email{carlos.bertulani@tamuc.edu}
\and A. Bonaccorso \at INFN, Sez. di Pisa, Largo B. Pontecorvo 3,
56127 Pisa, Italy. \email{angela.bonaccorso@df.unipi.it}}
%
%

\maketitle
\abstract{
  In this brief review we discuss the basic theoretical concepts used in the experimental studies of the most common cases of direct reactions such as (a) elastic scattering, (b) inelastic scattering, (c) Coulomb excitation, (d) transfer reactions and (e) breakup reactions.  }
\section{Introduction}
Direct nuclear reactions occur in a characteristic time  of $10^{-22}$ s, the time it takes a nucleon to cross the nucleus. The short time  only allows for the interaction with one or  a few nucleons on the surface of the nuclei involved. Contrary to compound nucleus products, the direct  reaction products are not distributed isotropically, but are focused in the forward direction. Bethe and Butler \cite{BB} first
understood the mechanism of direct reactions as due mainly to a surface diffractive effect and showed the way to use them as spectroscopic tools. A recommended list of references, which will help the reader of this short and scope-limited review are \cite{feshbach1992,Satchler:1983a,bwb,Book:Ber04,canto2013scattering}. The dawn of direct reaction theory and  its use to study nuclear structure has been recently reviewed  in \cite{me}.
\section{Elastic scattering} 
Upon hitting a target nucleus the wave function of an impinging nucleus is modified by the
scattering potential\footnote{For simplicity we assume a spherically symmetric potential.}  $U(r)$, leading to the
appearance of a phase-shift of the outgoing part of the scattered wave. {\it Elastic
scattering} occurs when there is no final energy transfer to the target nucleus.
Generally, the projectile wave function is not only modified by  a phase factor, but its magnitude might also change due to a loss of flux from the
elastic channel. For a projectile with momentum $p=\hbar k$, the total wavefunction at a distance $r$ from the scattering center  is asymptotically given by
\begin{equation}
\Psi \rightarrow {1\over2i}\sum^\infty_{\ell=0}(2\ell+1)i^lP_\ell(\hbox{cos}\theta){S_\ell e^{i(kr-\ell\pi/2)}
-e^{-i(kr-\ell\pi/2)} \over kr},
\label{Psiscatt} 
\end{equation} 
where $\theta$ is the scattering angle and a sum is carried out for all {\it partial waves} (or quantized angular momenta) $\ell = 0,1,2, \cdots$. The complex coefficient $S_\ell $  is known as {\it scattering matrix}, or simply $S$-matrix. It is called a matrix because when more than one reaction possibility (or {\it reaction channel}) is available, the complex coefficient $S_\ell $ may acquire multiple labels.  If
$S_{\ell}=1$, the sum in Eq. (\ref{Psiscatt}) leads to $\Psi \sim \exp(i{\bf k.r})$; that is, a plane wave. But if
\begin{equation}
 S_\ell=\exp[2i\delta_\ell]   ,
\end{equation} 
and $\delta_\ell$ is real, the incoming and
outgoing waves have the same magnitude and the scattering is termed
{\it elastic}. The quantities $\delta_l$  are known as {\it phase shifts}.

The {\it partial wave
expansion}, with the labels $\ell=0, 1, 2 ,\cdots = $
(s, p, d, $\cdots$ waves) includes each of the angular momentum components $\ell$  (in units of $\hbar$) of the scattered wave.
In classical mechanics, the angular momentum is given by
$\ell=kb$, where $b$ is the {\it impact parameter}, i.e.,  the
perpendicular distance to the target if the projectile would move along a straight-line.
But in quantum mechanics $\ell$ is not a continuous variable, varying in
steps of one. To determine the phase-shifts, and consequently the full scattering wave, one solves the Schr\"odinger equation (SE) for a given partial wave $\ell$ and its component $m$ along the incident direction. For a spherically symmetric potential, one can write $\Psi_{\ell m}=Y_{\ell m} (\hat{\bf r})u_\ell(r)/(kr)$  where $Y_{\ell m}$ is a spherical harmonics function. One obtains
\begin{equation}
    -{\hbar^2 \over 2 \mu}\left[{d^2 \over dr^2} - {\ell (\ell +1) \over r^2}\right] u_\ell (r) +U(r)u_\ell (r) = Eu_\ell (r),\label{partwav}
\end{equation}
where $\mu$ denotes the reduced mass of the system.
The numerical solution of this equation determines the modification of the partial wave $u_\ell$ with energy $E$ from an undisturbed partial wave component of the plane wave. By matching the solution at large distances with the asymptotic plane-wave component, the phase-shift is determined. For charged particles, the Coulomb potential $U_C$ leads to an analytical solution of Eq. (\ref{partwav}) and the phase-shift is obtained by matching the solution for $U=U_N +U_C$, with the asymptotic Coulomb wave, where we denote $U_N$ as the short-range part (nuclear) of the potential.

\begin{figure}[t]
\begin{center}
{\includegraphics[width=11cm]{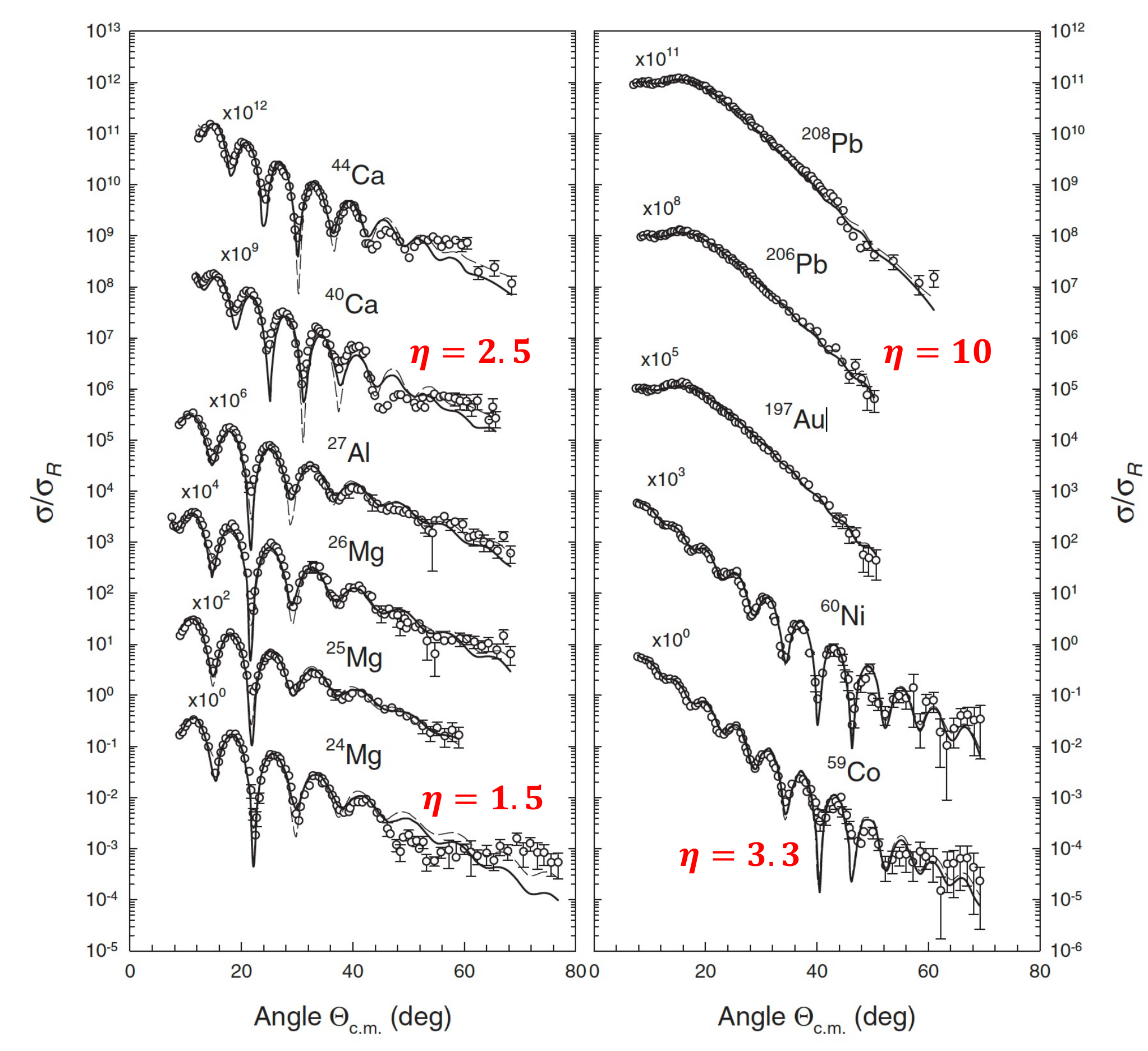}}
\end{center}
\caption{
Ratio between the measured cross sections to the Rutherford scattering cross sections for $^6$Li projectiles incident on different targets at 88 MeV. One clearly sees the transition from classical to quantum scattering as the Sommerfeld parameter $\eta$ decreases. Full quantum interference is visible for very small values of $\eta$ (Fraunhofer scattering). (Adapted from Ref. \cite{Hossain_2013}). }
\label{elast1}
\end{figure}

Adding all partial waves, Eq. (\ref{Psiscatt}) can be rewritten as a sum of a plane wave and a
scattering outgoing wave, $\Psi \sim \exp(i{\bf
k.r})+f(\theta) e^{ikr}/r$, with $f(\theta)$ accounting for the
distortion of its outgoing part at the scattering angle $\theta$. $f(\theta)$
is known as the {\it scattering amplitude}:
\begin{equation}
f(\theta)={1\over2ik}\sum_{
\ell=0}^\infty(2\ell+1)(S_\ell-1)P_\ell(\hbox{cos}\theta). \label{ftheta}
\end{equation}
The {\it differential scattering cross section} is obtained counting the number of
particles scattered through an angle $\theta$. This is achieved by calculating the particle current from the SE associated with the
wavefunction $\Psi$. One obtains
\begin{equation}
{d\sigma_e\over d\Omega}=\vert
f(\theta)\vert^2={1\over4k^2}\biggl\vert\sum^\infty_{\ell=0}(2\ell+1)(1-
S_\ell)P_\ell(\hbox{cos}\theta)\biggr\vert^2.
\label{sige}
\end{equation} 
The total scattering cross section is obtained by an integration over angles, yielding
\begin{equation}
\sigma_e=\pi
\lambdabar ^2\sum^\infty_{\ell=0}(2\ell+1)\vert1-S_\ell\vert^2,
\label{sige2}
\end{equation}
with $\lambdabar =\lambda/2\pi=1/k $. This method works for potentials that decay faster than $1/r$, or short-range potentials. For the Coulomb potential the    scattering amplitude becomes
\begin{equation}
f_C(\theta) = - {\eta \over 2k\sin^2(\theta/2)} e^{-i\eta\ln(\sin^2\theta/2)}e^{2i\sigma_0},
\end{equation}
where $\eta=Z_1Z_2e^2/\hbar v$ is the {\it Sommerfeld parameter}, with $Z_1$ and $Z_2$ being the nuclear charges, $v$ their relative velocity, and $\sigma_0=\arg\Gamma(1+i\eta)$ is the {\it Coulomb phase}. 
When the scattering occurs under the influence of  short-range plus Coulomb potentials the correct amplitude entering Eq. (\ref{sige}) is
\begin{equation}
    f(\theta) = f_C(\theta) + f_N(\theta),
\end{equation}
where $f_N$ is the amplitude due to the short range potential only.

In the presence of more than one-channel, the elastic scattering may be influenced by the coupling between them, requiring the solution of coupled  differential equations involving interactions between the channels, after which the $S$-matrices can be computed and the prescription leading to the Eq. (\ref{ftheta}) can be used. Some of the channels can be {\it inelastic}, absorbing energy and thus influencing the elastic scattering channel. Often, the number of channels are too many to be treated individually, and one introduces the concept of a complex {\it Optical Potential} (OP), $U_{OP}$, the imaginary part of which being responsible for the absorption into the inelastic channels. The elastic cross section is still given by Eq. (\ref{sige2}), but the magnitude of the $S$-matrices for the elastic channel becomes smaller than unity. The absorption, or {\it reaction cross section}, $\sigma_r$, is given by 
\begin{equation}
\sigma_r=\pi
\lambdabar ^2\sum^\infty_{\ell=0}(2\ell+1)\left[1-|S_\ell\vert^2\right].
\label{siger}
\end{equation}
This equation has a simple interpretation. $\pi \lambdabar ^2$ is the ``quantum area'' for a projectile with momentum $k=1/\lambdabar$, $2\ell +1$ is the number of magnetic states for an angular momentum $\ell$, and $ 1-|S_\ell\vert^2$ is the absorption probability for the partial wave $\ell$ 

In a different formalism,  the scattering amplitude can be written as
\begin{equation}
    f(\theta) = -{\mu \over 2 \pi \hbar^2}\left< {\bf k}' |U | \psi^{(+)}_{\bf k} \right> =-{\mu \over 2 \pi \hbar^2} T({\bf k'},{\bf k}),
\end{equation}
where $\left|{\bf k}'\right>$ is a Dirac ``ket'' notation for a plane wave with momentum ${\bf k}'$ and $\left|\psi^{(+)}_{\bf k}\right>$ is the outgoing wave with momentum ${\bf k}$. $T({\bf k'},{\bf k})$ is known as the {\it transition-, or T-matrix,} where 
\begin{equation}
    \left< {\bf k}' |U | \psi^{(+)}_{\bf k} \right> = \left< {\bf k}' |T | {\bf k} \right> = T({\bf k'},{\bf k}). \label{Top}
\end{equation}
A formal solution of the scattering problem requires the knowledge of the T-matrix. This can be achieved by iteration of the full Schr\"odinger equation, writing it in an operator form so that $H\psi = E\psi$. For positive energies $E$, $\psi$ is the scattering wavefunction. $H = H_0 + U$ is the Hamiltonian, with $H_0$ being here the kinetic energy operator $-\hbar^2\nabla^2/2\mu$. The formal solution for the T-matrix, entering Eq. (\ref{Top}) is 
\begin{equation}
    T=U+UG_0T= U + UG_0U + \cdots\label{lipp},
\end{equation}
known as the {\it Lippmann-Schwinger (LS) equation}, where the so-called {\it Green's function operator} is defined as $G_0=(H_0-E)^{-1}$. The LS equation can be solved by iteration, as is indicated by the r.h.s of (\ref{lipp}). 

At high energies, an approximate solution of the Schr\"odinger equation yields the {\it eikonal phase-shift} \cite{glauber1959lectures} 
\begin{equation}
2\delta(b)=-{1\over \hbar v}\int U_{OP}(r) dz,
\label{deltab}
\end{equation}
where the radial coordinate describing the projectile-target relative motion is split into a transverse, ${\bf b}$, and a longitudinal component, $z$ so that ${\bf r} \equiv ({\bf b},z)$. The coordinate ${\bf b}$ is often interpreted as the impact parameter variable in classical mechanics. In the high energy regime, the sum over partial waves in Eq. (\ref{ftheta}) involves too many terms. One can approximate the sum by an integral over $b$ using  $\ell=kb$ as a continuous variable. This procedure yields the simple formula \cite{glauber1959lectures} 
\begin{equation}
f(\theta) = ik\int db\, b \, J_0(qb)\left[ 1- S(b)\right],
\label{eik}
\end{equation}
where $q=2k\sin(\theta/2)$ is the {\it momentum transfer} in the collision, with the $S$-matrix  given by $S(b) = \exp[2i\delta(b)]$ in the eikonal form. Again, for {\it multi-channel reactions} one needs to solve  coupled channels equations.

At low energies, and/or in reactions involving highly charged nuclei, the elastic scattering is predominantly a {\it Coulomb scattering}. The differential scattering cross section is given by the {\it Rutherford formula}, $d\sigma_R/d\theta = a^4 /\sin^4 (\theta/2)$ with $a=Z_1Z_2 e^2/\mu v^2$.  A rough visualization of the scattering process assumes that  the incoming wave splits into two pieces, one passing by one side and the other passing by the opposite side of the target (conveniently called the {\it near-} and {\it far-sides}). When these two pieces do not interfere, one recovers the {\it classical scattering}.  A good measure of the passage from the classical  to quantum scattering regime is obtained by using the {\it Sommerfeld parameter} $\eta = Z_1Z_2e^2/\hbar v$. When $\eta$ decreases,  the scattering is increasingly influenced by the interference of the near- and the far-side waves. Around $\eta \sim 10$ the scattering changes from the so-called {\it Fresnel} to the {\it Fraunhofer} regime.  This is clearly seen in Figure \ref{elast1}.

It is clearly visible on the left panel of Figure \ref{elast1} that the elastic scattering in the Fraunhofer regime displays wiggles that are nearly equally displaced. The distance between the dips in the angular distribution are telltales of the size $R$ of the system involved, i.e., $\Delta \theta \sim 1/kR$, whereas the nearly exponential fall-off of the cross section is due to the diffuseness $a$ of the nuclear surface, i.e., $\sigma(\theta) \propto \exp(-qa)$, with $q$ being the momentum transfer defined previously. Therefore, elastic scattering is a very good probe of the {\it nuclear geometry} of the optical potential.

\begin{figure}[h!]
\centering
\includegraphics[scale=1.5]{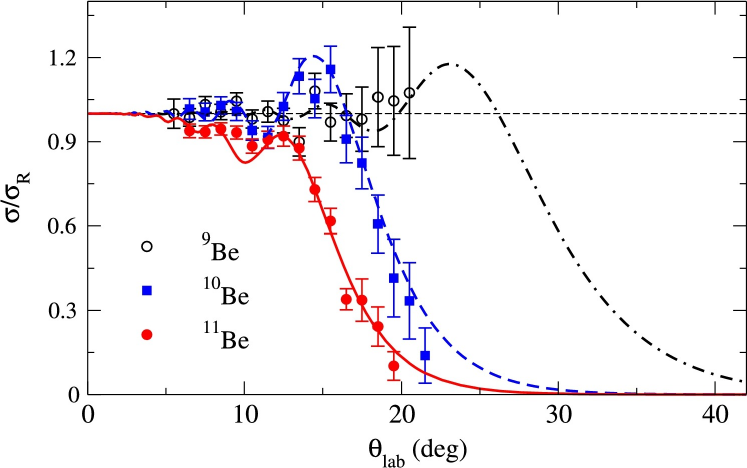}
\caption {Examples of elastic scattering angular distributions from Ref. \cite{2020135942}. }
\label{figang}
\end{figure}

\section{Inelastic scattering}
{\it Inelastic scattering} occurs when a state in the projectile or target nucleus is excited in the reaction. Consider a nucleus in an initial state $|\left. i\right>$ as an eigenstate of the internal nuclear Hamiltonian $H_0$. In a direct reaction the nucleus will be subject to an interaction potential $U$ with another nucleus (or with a nucleon). If this interaction can be treated perturbatively, {\it Fermi's Golden rule} can be used to obtain the cross section for the excitation a final state $|\left. f\right>$, i.e.,
\begin{equation}
\sigma = {\mu \over \hbar k}{2\pi \over \hbar}\left| \left< f |U|i \right>\right|^2 \rho(E_f) ,
  \label{FGR}
\end{equation}
where  $\rho$ is the {\it density of the final states}, i.e., the number of states per energy interval at the final energy $E_f$. The nuclear wavefunctions  $\left| i,f\right>$ include the internal wavefunction $\left| \phi\right>$ of the excited nucleus as well  as  the relative motion wavefunction, $|\left.\chi\right>$.  

\begin{figure}[t]
\begin{center}
{\includegraphics[width=7cm]{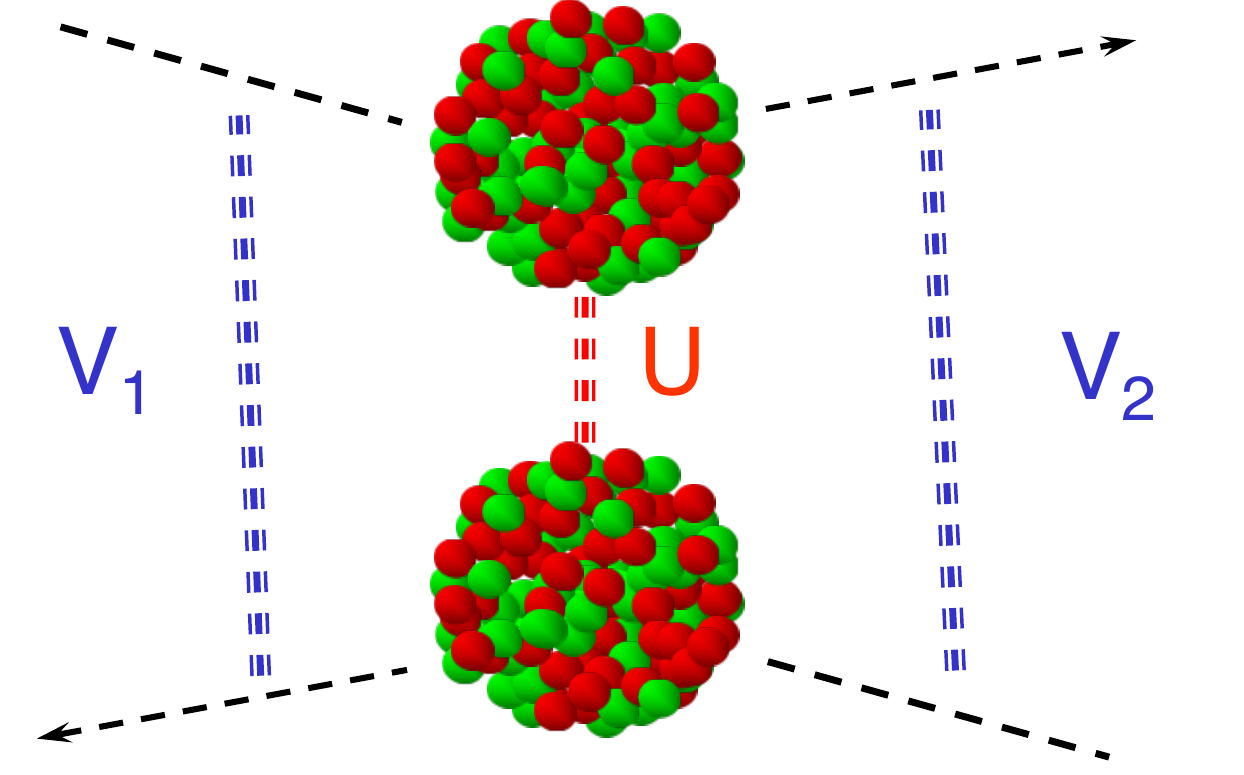}}
\end{center}
\caption{
Schematic representation of inelastic scattering described within DWBA. The nuclear interaction is taken to first order (the B in DWBA) and the distorted waves are calculated exactly given the entrance, $V_1$, and outgoing channel, $V_2$, optical potentials.} 
\label{dwba}
\end{figure}

The ``bra'' and ``ket'' notation used in the above relation can be rewritten in terms of the {\it intrinsic nuclear coordinate}, ${\bf x}$, and the {\it relative motion coordinate} ${\bf y}$,
\begin{equation}
\sigma  = 
{2\pi \mu\over \hbar^2 k} \rho(E_f) \left| \int d^3 x d^3y \phi^*_f({\bf x}) \chi^{(-)*}({\bf y}) U({\bf x,y}) \phi_i({\bf x}) \chi^{(+)}({\bf y}) \right|^2,
  \label{FGR2}
\end{equation}
where the notation $\chi^{\pm}$ stands for outgoing (incoming) relative motion wavefunction. In the {\it Plane Wave Born Approximation} (PWBA), the wavefunctions $\chi$ are taken as plane waves and the integral above becomes a Fourier transform for the momentum transfer ${\bf q} = {\bf k}' - {\bf k}$. But, the relative motion wavefunction $\chi$ can be calculated exactly by solving the SE for the scattering part separately, with the proper energies $E_i$ and $E_f$ and optical potentials $V_1$ and $V_2$ for the incoming and outgoing channel, respectively.  In this case, Eq. (\ref{FGR2}) is known as the {\it Distorted Wave Born Approximation} (DWBA).   Figure \ref{dwba} shows schematically that  the terminology ``Born'' in the DWBA means that the perturbation theory is treated to first-order only, i.e., the excited nucleus is assumed to interact only  once and weakly with the other nucleus, while ``distorted'' means that  the relative motion between the nuclei in the entrance and the outgoing channels is taken into account to all orders.

Angular distributions for inelastic scattering display similar features as elastic scattering such as oscillations, with the distance between dips becoming a telltale of the geometry of the reacting nuclei. However, in contrast to elastic scattering, and due to the presence of the interaction potential $U$ in Eq. (\ref{FGR2}), inelastic scattering data do not often display an exponential fall off at the Fraunhoffer regime because of the diffuseness of the nuclear densities. The details of the inelastic cross section will contain information of $U({\bf r})$, as well as the internal wavefunctions $\phi$.  If the optical potential (determining the distorted waves) and the interaction potential (responsible for the excitation) are both well known, inelastic scattering can be used as a tool for spectroscopic studies of the internal wavefunction $\phi$. 

\subsection{Coupled channels}
The Schr\"odinger equation (SE) of a nuclear system subject to an external agent $V(t)$ is governed by the Hamiltonian $H(t)=H_{0}+V(t)$ where $H_0$ is the non-perturbed Hamiltonian $H_{0}$. Let us assume  that $H_{0}$ has eigenvalue and eigenfunction solutions, i.e., that $ H_{0}\psi_{n}({\bf r})=E_{n}\psi_{n}({\bf r})$, and that the eigenfunctions $\psi_n$ form a complete basis. The total wavefunction $\Psi$, obeying the time-dependent SE, $ H\Psi=i\hbar\partial \Psi/\partial t$, may be expanded as $\Psi=\sum_{n}a_{n}(t)\psi_{n}e^{-iE_{n}t/\hbar}$, where $a_n$ are time-dependent coefficients. Inserting this expansion into the time-dependent SE leads to
\begin{equation}
i\hbar\sum_{n}\dot{a}_{n}\psi_{n}e^{-iE_{n}t/\hbar}=\sum_{n}Va_{n}\psi
_{n}e^{-iE_{n}t/\hbar},\label{5.102}%
\end{equation}
with $\dot{a}_{n}\equiv d a_{n}(t)/d t$. From the
orthogonalization properties of the $\psi_{n}$,  multiplying (\ref{5.102})
by $\psi_{k}^{\ast}$ and integrating it over the ${\bf r}$, one obtains 
the {\it time-dependent coupled-channels} equations%
\begin{equation}
\dot{a}_{k}\left(  t\right)  =-\frac{i}{\hbar}\sum_{n}a_{n}\left(  t\right)
\ V_{kn}\left(  t\right)  \ e^{i{\frac{E_{k}-E_{n}}{\hbar}}t},\label{5.103}%
\end{equation}
where  $V_{kn}=\int\psi_{k}^{\ast}V\psi_{n}\,d^3r$.

A simple derivation of the {\it time-independent coupled-channels equations} can be achieved for high energy collisions. The replacement $z=vt$ can be done for an almost undisturbed trajectory of a projectile with velocity $v$ passing by the target nucleus with an impact parameter $b$. The general coupled equations are then simply an outcome of Eq. (\ref{5.103}), i.e., \cite{BertulaniPRL94.072701} 
\begin{equation}
{i\hbar v}\frac{d}{d z}{\cal A}_c(b,z) = \sum_{c'}\left\langle
\Psi_{c}|V_{int}(b,z)|\Psi_{c'}\right\rangle
{\cal A}_{c'}(b,z)\ e^{iE_{cc'}  z/v}, \label{cceq4}
\end{equation}
where we introduced the channel index $c=\ $\{$i$, $\ell$, $m$\}; with $i$ denoting one of the nuclear states $i>0$, $i=0$ the ground state, and $\ell$ and $m$\ are the orbital angular momentum and its projection along the incident $z$-axis. $E_{cc'}=E_{c'}-E_c$ is the excitation energy. The amplitudes $a_c$ were renamed to ${\cal A}_{c}(b,z)$. By solving these equations, using the initial condition,
${\cal A}_c(b,-\infty) = \delta_{c0}, $
one can obtain the probability that a channel $c$ is populated in the reaction, $|{\cal A}_{c}(b,\infty)|^2$.

\begin{figure}[t]
\begin{center}
{\includegraphics[width=12cm]{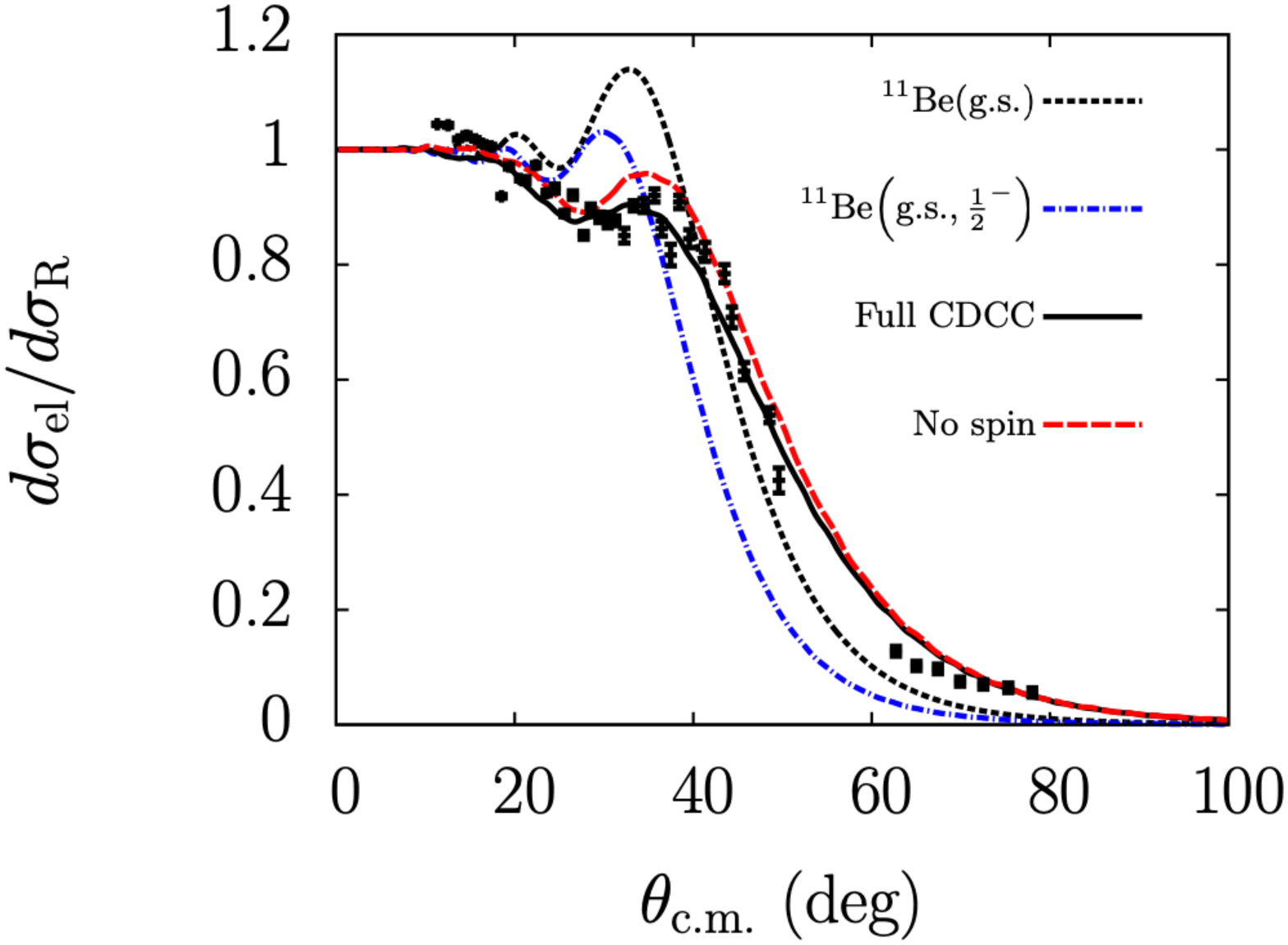}}
\end{center}
\caption{CDCC calculations \cite{Druet2012} for $^{11}$Be + $^{64}$Zn at 24.5 MeV compared to experimental data  \cite{PhysRevLett.105.022701}.
$^{11}$Be(g.s.) and $^{11}$Be(g.s., $1/2^-$) denote the one-channel (ground state only) and two-channel calculations (ground state and the first excited state).
The elastic cross sections are divided by the Rutherford cross sections. Calculations based on  CDCC, shown as a continuous curve nearly match the data, taking into account a discretized continuum. The long-dashed curve neglects the spin of the projectile.}
\label{cdcc}
\end{figure}

At lower energies, where the partial-wave expansion is more adequate, one can easily deduce the corresponding coupled-channels equations from Eq. (\ref{cceq4})  by using the correspondence of the continuous variable $b$ with the angular momentum\footnote{To be more precise, one should use $b\rightarrow (\ell+1/2)/k$.}: $b \rightarrow \ell/k$. The integrals over the impact parameter become a sum over the partial waves $\ell$. Finally, the consideration of total angular momenta and spin coupling can be accounted for, with channels discerning by the angular momentum quantum numbers ($J,M$). 
 
At high energies,  the angular distribution of the inelastically scattered particles for the excitation of the channel state $c$ is obtained from
\begin{equation}
f_{c}^\mu(\theta)=ik\int_0^\infty db b J_\mu(qb){ S}(b){\cal A}_c(b, \infty), \label{cceq6}
\end{equation}
with $k$ being the projectile wavenumber, and $q=2k \sin(\theta/2)$ the momentum transfer. Here, we simplified the notation  using $\mu = M_c - M_0$, with $M_i$ being the magnetic quantum number associated with the total angular momentum $J_i$. Averaging over the initial spin and summing over the final spin, yields the differential cross section
\begin{equation}
{d\sigma_{c} \over d \Omega} = {1\over 2J_0+1} \sum_{M_0,M_{c}}\left| f_{c}^{(M_c-M_0)}(\theta) \right|^2. \label{cceq7}
\end{equation}
At high energies, the S-matrices in Eq. (\ref{cceq6}) are given by ${ S}(b)=\exp\{2i\delta(b)\}$ where the phase-shift can be related to the corresponding nucleon-nucleon (NN) scattering quantities \cite{HUSSEIN1991279}. In this case,  the eikonal phase becomes
\begin{equation}
2\delta(b)= \int {\rho}_{P}(q){\Gamma}(q){\rho}_{T}(q) J_0(qb)q \,dq ,\label{eik7}
\end{equation} 
where $J_{0}$ is the ordinary Bessel function of zeroth-order, and the nucleon-nucleon scattering profile function is parametrized as \cite{Ray:1979}
\begin{equation}
{\Gamma}(q)=\frac{i +\alpha_{NN}}{4\pi }\sigma_{NN}e^{-\beta_{NN}q^2} .\label{eik8}
\end{equation}
$\sigma_{NN}$ is the total nucleon-nucleon cross section,  $\alpha_{NN}$  is the ratio between the real and the imaginary part of the $NN$-scattering amplitude. $\beta_{NN}$ is a momentum dependence parameter. These parameters are fitted to reproduce the NN scattering observables. Tables with the energy dependence of these parameters are given in Refs. \cite{HUSSEIN1991279,AUMANN2020103753}.

The total excitation cross section for channel $c$ averaged over  the initial spin $J_0$ is given by
\begin{equation}
\sigma_{c} = {2\pi \over 2J_0 +1} \int db \, b |{S}(b)|^2\left|{\cal A}_c(b,\infty)\right|^2 .\label{cceq8}
\end{equation}
The coupled-channels method can be used with any nuclear structure model, either  from  a two-body, three-body, or many-body description of the nucleus enabling the calculation of the matrix elements $\left\langle \Phi_{c}|V_{int}(b,z)|\Phi_{c'}\right\rangle$ in Eq. (\ref{cceq4}). {\it First order excitation amplitudes} can be calculated from Eq. (\ref{cceq4}) replacing ${\cal A}_c(b,z)=\delta_{c0}$ on its right-hand side, leading to
\begin{equation}
{\cal A}_c(b,\infty) = -{i\over \hbar v}\int_{-\infty}^\infty dz \left\langle
\Phi_{c}|V_{int}(b,z)|\Phi_0\right\rangle
\ e^{iE_{0c}  z/v}. \label{cceq9}
\end{equation}
If the states $c$ are in the continuum, then Eq.  (\ref{cceq7}) means
\begin{equation}
{d\sigma_c \over d \Omega dE} = {1\over 2J_0+1} \sum_{M_0,M_{c}}\Big\vert f^{(Mc-M_0)}(\theta, E=E_c) \Big\vert^2, \label{cceq10}
\end{equation}
for one of the states in the continuum  with energy $E_c$.

The angular distribution integrated over all continuum energies is given by
\begin{equation}
{d\sigma_c \over d \Omega } = {1\over 2J_0+1} \sum_{M_0,M_{c}}\int dE \, \Big\vert f^{(Mc-M_0)}(\theta, E) \Big\vert^2. \label{cceq11}
\end{equation}
For large bombarding energies so that $q\simeq k \theta\simeq k\sin\theta$, it follows that $d\Omega=2\pi q dq/k^2$, and
\begin{eqnarray}
\sigma_c &=& {2\pi\over 2J_0+1} \sum_{M_0,M_{c}}\int db b \int db' b'  \int dq  q J_\mu(qb) J_\mu(qb') {S} (b) {S}^*(b') \nonumber \\
&\times& {\cal A} (b,\infty) {\cal A}^*(b',\infty) = {1\over 2J_0+1} \int db \, b |{S}(b)|^2\left|{\cal A}_c(b.\infty)\right|^2. \label{cceq12}
\end{eqnarray}

In calculations involving weakly-bound nuclei the transition to the continuum (breakup) is followed by a coupling between states in the continuum. The states in the continuum can be treated as isolated {\it discretized states}. Such formalism is known as the {\it continuum-discretized coupled-channels} (CDCC) equations.    
Figure \ref{cdcc} shows an example of a coupled-channels calculation with the addition of couplings in the continuum \cite{Druet2012}. One sees that the inclusion of channel coupling considerably improves the comparison with the experiment   \cite{PhysRevLett.105.022701}. Due to the relatively low energy of the projectile (24.5 MeV), the CDCC calculations are  in this case performed using the partial wave expansion method.

\subsection{The optical potential}

The simple description of high-energy collisions presented above is not adequate at lower energies and a good knowledge of an {\it optical potential} $U$ will be the most important ingredient in the calculations of elastic and inelastic scattering. The optical potential (OP) contains information about the leakage of probability from a channel of interest, e.g., the elastic channel, to other channels involving the excitation of nuclear states. Traditional theoretical methods assume that all our ignorance about the numerous relevant channels can be simulated by the introduction of an imaginary potential $W$. A popular form of the  {\it phenomenological OP} is
\begin{equation}
U(r)=V_0(r) + V_s(r)({\bf l}\cdot {\bf s}) + V_C(r) + iW(r), \label{Wspot}
\end{equation}
where $V_0$ is the {\it central part} of a real potential, $V_s$ is a {\it spin-orbit component}, and $V_C$ is the {\it Coulomb potential} between the nuclei. The functional forms of the potentials $V_0$ and $W$ are usually taken as a {\it Woods-Saxon (or Fermi) function} 
\begin{equation}
    f(r) = {f_0 \over 1+\exp\left({r-R_0\over a_0}\right)}, \label{fr}
\end{equation}
where $f_0$ ({\it strength}), $R_0$ ({\it radius}) and $a_0$ ({\it diffuseness}) are taken as adjustable parameters. The second term on the r.h.s. of Eq. (\ref{Wspot}) is usually taken as a derivative of the function (\ref{fr}) and accounts for the increased probability of nucleon-nucleon collisions at the nuclear surface due to a decrease of the {\it Pauli principle} at lower nucleon densities.  It also accounts for the surface preference of the spin-orbit potential.

Many other methods exist  to deduce the optical potential from  basic details of nuclear structure and reactions. A popular case is the {\it folding potential}, obtained from the ground-state nuclear densities $\rho ({\bf r})$ and the nucleon-nucleon potential
\begin{eqnarray}
U({\bf r}) &=& \int d^3 r' v_{NN}({\bf r}'-{\bf r}) \rho_A ({\bf r}'), \ \ \ {\rm for \ nucleon-nucleus},\nonumber \\
&=&\int d^3 r' v_{NN}({\bf r}'+{\bf r}''-{\bf r}) \rho_A ({\bf r}') \rho_B ({\bf r}''). \ \ \ {\rm for \ nucleus-nucleus}. \label{vnnmed}
\end{eqnarray}
To these real potentials, corresponding imaginary parts are introduced usually with the same form as the real parts multiplied by constants adjusted to reproduce experimental observables. A modification of these equations can be introduced to account for {\it medium effects} of the nucleon-nucleon interaction, $v_{NN}$, inducing an energy dependence $v_{NN}({\bf r}, E)$ to obtain  a corresponding energy-dependent optical potential $U({\bf r},E)$.  A variety of {\it effective interactions} accounting for medium effects have been developed such as the M3Y \cite{BERTSCH1977399},  Love-Franey \cite{PhysRevC.31.488}, the JLM  \cite{PhysRevC.16.80}, and many other popular optical potentials.

A {\it microscopic formalism} for the optical potential can be developed by linking the T-matrix for the nucleus-nucleus collisions to  T-matrices for nucleon-nucleon scattering. As a starting point one uses the Lippmann-Schwinger equation for the whole system of $A+B$ nucleons, yielding the approximate equation \cite{Feshbach:1958}
\begin{equation}
    U=\sum_{i\neq j}t_{ij} +\sum_{i\neq j}t_{ij} {1 - |\Psi_0><\Psi_0| \over E+\hbar^2 \nabla^2/2\mu +i\epsilon}t_{ij} + \cdots = U_{LO} + U_{NLO} + \cdots , \label {Uopt}
\end{equation}
where the sum runs over all nucleons, $E$ is the total energy of the colliding nucleus-nucleus system, $t_{ij}$ are the T-matrices for (free) nucleon-nucleon scattering for their relative motion, $\nabla$ is a derivative of the center-of-mass motion of the system, and $\left|\Psi_0\right>=\left|\Psi^{A}_0\right>\left|\Psi^{B}_0\right>$ is the product of the ground-state wavefunctions for nucleus $A$ and $B$. The factor $i\epsilon$ is included in Eq. (\ref{Uopt}) to account for the proper outgoing wave boundary condition. As shown in Ref. \cite{HUSSEIN1991279}, this formalism allows for an understanding on how the optical potential is influenced by multiple nucleon-nucleon collisions, where $U_{LO}$ is the {\it leading-order} optical potential, $U_{NLO}$ the {\it next-to-leading order} correction, and so on. For heavy nuclei, $U_{NLO}$ and higher-order corrections still comprise an important part of the whole optical potential  \cite{HUSSEIN1991279}. 

The medium effects on the nucleon-nucleon scattering cannot be neglected in most cases. Introducing the Pauli-principle for nucleon-nucleon scattering in the medium implies solving an equation similar to the Lippmann-Schwinger equation, known as the {\it G-matrix}, and often written the form of the {\it Bethe-Goldstone equation} \cite{Gomes1958} 
\begin{equation}
 \left<{\bf k}'|G|{\bf k}\right>   =\left<{\bf k}'|v_{NN}|{\bf k}\right> + \int {d^3 k'' \over (2\pi)^3} {\left<{\bf k}'|v_{NN}|{\bf k}''\right>Q({\bf k}'')\left<{\bf k}''|G|{\bf k}\right> \over E({\bf P}'',{\bf k}'') -E_0 + i\epsilon} ,
\end{equation}
where $E({\bf P}, {\bf k}) = e({\bf P} + {\bf k})+ e({\bf P} - {\bf k})$ are off-shell nucleon single-particle energies, with $\bf P$ being the nucleon-nucleon center of mass momentum and $\bf k$ their relative momentum. $E_0$ is the on-shell energy, i.e., when the final and initial  energies and momenta  of the nucleons obey conservation laws. 
As with the case of the Lippmann-Schwinger equation, the Bethe-Goldstone equation can be solved by iteration. The operator $Q({\bf k}'')$ is equal to one if the individual momenta of the nucleons are larger than the local Fermi momentum, i.e., if $|{\bf k}_{1,2}| > k_F$, where ${\bf k}_{1,2} = {\bf P} \pm {\bf k}$. If this condition is not valid,   $Q({\bf k}'') = 0$.

A similar method accounts for the medium modification of the nucleon-nucleon force and is known as the {\it Brueckner theory}. Similarly to the Bethe-Goldstone equation, the nucleon-nucleon interaction in the nuclear medium is formally given by \cite{Goldstone267}  
\begin{equation}
   v(p) =  \left< p|v|p\right> = Re \sum_{k<k_F} \left< pq | G | pq-qp\right>,
\end{equation}
where $\left|pq-qp\right>$ is a short-hand notation of the anti-symmetrization of the nucleon wavefunctions. In the Brueckner scheme this equation is solved self-consistently, i.e., the single-particle energies $e$ depend on the nucleon-nucleon potential $v(p)$, which depends on the solution for the G-matrix, which in turn depends on $e$ and $v(p)$. 

A microscopic description of the optical potential has to include the loss of energy into nuclear excitation and the details of the nuclear wavefunctions. A traditional method to tackle these features uses the concept of {\it self-energies} induced by all possible intermediate states. For example, in the particle-vibrator coupling model \cite{BERNARD198075} the optical potential arises from the relation
\begin{eqnarray}
U(E,{\bf r},{\bf r}') = U_{HF}({\bf r},{\bf r}')  + \Sigma(E,{\bf r},{\bf r}'),
\end{eqnarray}
where $U_{HF}$ is a {\it mean field potential} and the self-energy is given by
\begin{eqnarray}
\Sigma(E,{\bf r},{\bf r}') = {1\over 2J_0+1}\left( \sum_{nL,p>F} {\left|\left<i||V||p,nL\right>\right|^2 \over E-\epsilon_p-E_{nL}-i\eta} +
\sum_{nL,h<F} {\left|\left<i||V||h,nL\right>\right|^2 \over E-\epsilon_h+E_{nL}-i\eta}\right),
\end{eqnarray}
with $|nL>$ denoting phonon states, $\left|p\right>$ ($\left|h\right>$) are particle (hole) states and $\epsilon_p$ ($\epsilon_h$) their respective energies. A nuclear structure model is used to obtain the wavefunctions corresponding to all the states entering this equation.

There exist other microscopic formalisms for the optical potentials such as the {\it dispersive optical potential model} (DOM) based on its functional analytical properties \cite{MAHAUX1991253}. It recalls the physical concept that a scattered  wave is only emitted after the arrival of the incident wave.  As a consequence, one finds that
\begin{eqnarray}
U(E,{\bf r},{\bf r}') = U_0({\bf r},{\bf r}') + V(E,{\bf r},{\bf r}') + iW(E,{\bf r},{\bf r}'),
\end{eqnarray}
where the real and imaginary part of the potential are related by the {\it dispersion relation}
\begin{eqnarray}
W(E,{\bf r},{\bf r}') = i{{\cal P}\over \pi} \int dE' {V(E,{\bf r},{\bf r}')\over E-E'},
\end{eqnarray}
where ${\cal P}$ denotes the principal value of the integral.  The DOM has been applied to a large number of data (see, e.g., Ref. \cite{PhysRevC.83.064605}) and microscopic {\it ab initio calculations} for the nuclear states have been incorporated in its numerical derivation \cite{PhysRevC.84.034616}. 

\section{Coulomb excitation}

{\it Coulomb excitation} is a specific inelastic scattering process where one
nucleus excites another via its electromagnetic (EM) field $V$.
The EM field can be decomposed in  a sum of multipoles,
for example, $E1$, $E2$, $M1$, $\cdots$, each one containing angular
momentum and parity. At low collision  energies,  $E2$ (electric quadrupole)
excitations are stronger, whereas at higher energies $E1$ (electric dipole)
excitations prevail. 

At low energies (below the Coulomb barrier), Coulomb excitation  has been used to analyze experiments on multiple excitations and reorientation effects \cite{AlderWinther1965,cline:1986:ARNPS}. At relativistic energies, the kinematics is characterized by near straight-line trajectories and by retardation effects due to special relativity  \cite{alder-winther}.  A full quantum mechanical theory of relativistic Coulomb excitation was developed in Refs. \cite{bertulani:1985:NPA,BERTULANI1993158}, including diffraction and absorptive effects. At intermediate energies ($20 < E_{lab} < 200$ MeV/nucleon), both retardation and relativistic corrections of the Rutherford trajectories are necessary for an accurate description of the reaction \cite{aleixo:1989:NPA}.

In first-order perturbation theory, the differential cross section  is given by 
\begin{eqnarray}
{\frac{d\sigma_{i\rightarrow f}}{d\Omega}}&=&
\frac{d\sigma_R}{d\Omega }\frac{16\pi^{2}Z_{2}^{2}e^{2}}{\hbar^{2}}
\sum
_{\pi\lambda\mu}{\frac{B(\pi\lambda,I_{i}\rightarrow
I_{f})}{(2\lambda +1)^{3}}}\mid
S(\pi\lambda,\mu)\mid^{2},\label{cross_2}
\end{eqnarray}
where $d\sigma_R/d\Omega$ is the Rutherford cross section and
$Z_2$ is the projectile charge.
$B(\pi\lambda,I_{i}\rightarrow I_{f})$ is known as the {\it reduced matrix element} of the excited nucleus, where $\pi\lambda=E1,\ E2,$
$M1,\ldots$ is the excitation multipolarity, and
$\mu=-\lambda,-\lambda+1,\ldots,\lambda$.
The {\it orbital integrals} $S(\pi\lambda,\mu$) include information on
the reaction dynamics, i.e., on the details of the EM fields \cite{aleixo:1989:NPA}.

Coulomb excitation is an external process,  occurring
when the nuclear matter of the nuclei do not    overlap. This implies that the
Coulomb excitation matrix elements  display the   
same form, or operators, as those for the excitation by real photons (except for $E0$ 
excitations, usually very small). As a consequence,  the  
Coulomb excitation cross sections can always be written as \cite{BERTULANI1988299} 
\begin{eqnarray}
\frac{d\sigma _{C}\left( E_{x}\right) }{dE_{x}}&=&\sum_{E\lambda }\frac{%
n_{E\lambda }\left( E_{x}\right) }{E_{x}}\sigma _{E\lambda }^{\gamma }\left(
E_{x}\right) +
\sum_{M\lambda }\frac{n_{M\lambda }\left( E_{x}\right) }{E_{x}}%
\sigma _{M\lambda }^{\gamma }\left( E_{x}\right) \;,  \label{sigmac}
\end{eqnarray}
where $\sigma {_{\pi \lambda }^{\gamma }}$ $\left( E_{x}\right) \;$are 
{\it cross sections induced by real photons}  (photonuclear cross sections) with multipolarity $\pi \lambda $. $E_{x}$ is the excitation (or photon) energy, and
$n_{M/E\lambda} (E_x)$ are known as {\it virtual photon numbers} \cite{BERTULANI1988299}.

\begin{figure}[t]
\begin{center}
{\includegraphics[width=10cm]{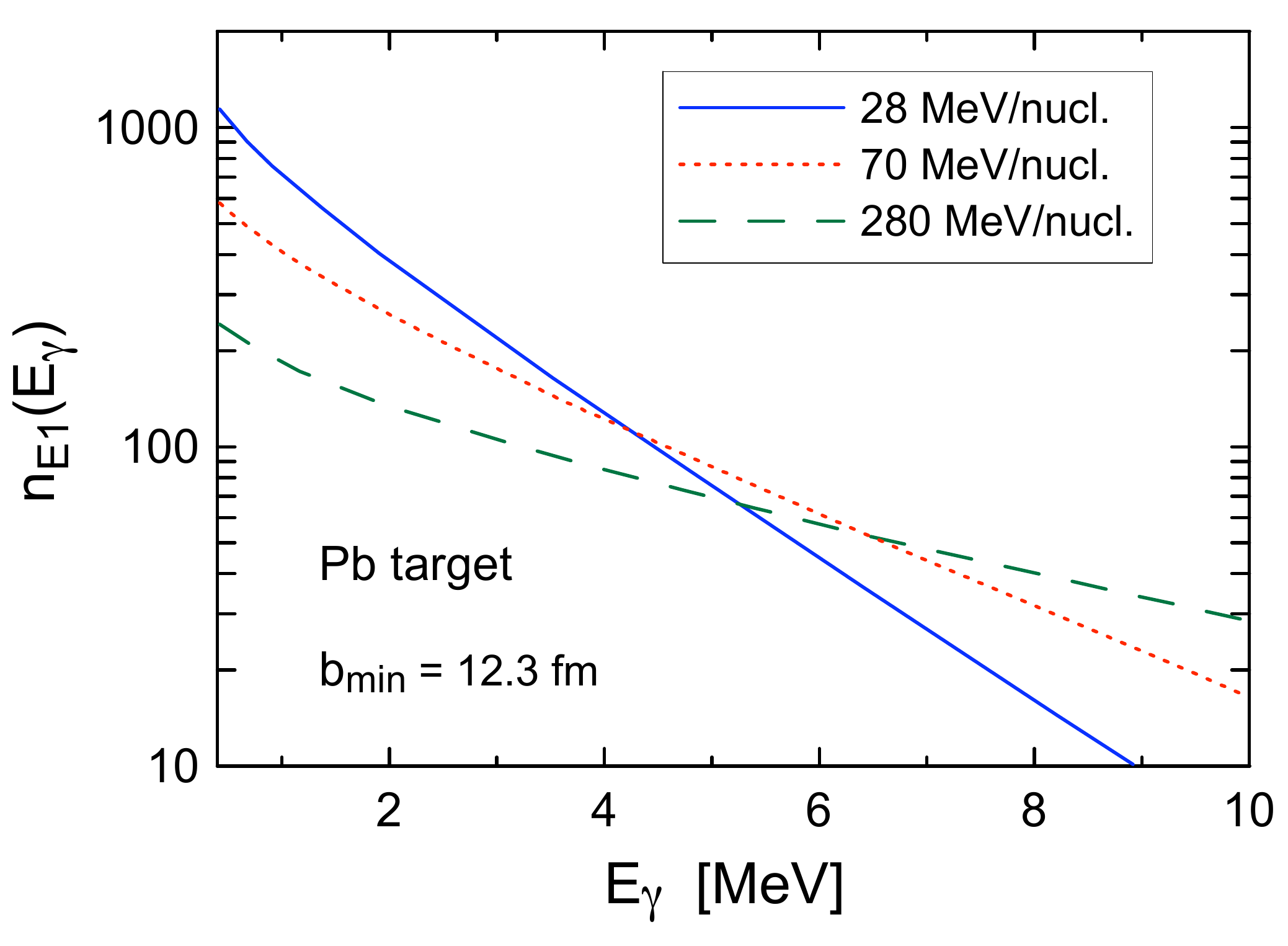}}
\end{center}
\caption{
Number of virtual photons with $E1$ multipolarity,  for three typical bombarding energies, ``as viewed'' by a Pb
projectile nucleus incident on a Pb target  with impact parameters larger than $b_{min}=12.3$ fm.} 
\label{vpn}
\end{figure}

{\it Photoabsorption cross sections} are functions of the reduced matrix elements,
for the excitation energy $E_{x}$. Explicity \cite{BERTULANI1988299}, 
\begin{equation}
\sigma _{\gamma }^{\pi
\lambda }(E_{x})=\frac{(2\pi )^{3}(\lambda +1)%
}{\lambda \left[ (2\lambda +1)!!\right] ^{2}}\left( \frac{E_{x}}{\hbar c}%
\right) ^{2\lambda -1}\frac{dB\left( \pi \lambda, E_{x}\right)}{dE_{x}},  \label{(1.2)}
\end{equation}
where $dB/dE_{x}$ are known as  {\it electromagnetic response
functions}. The {\it total transition strength} is the integral 
\begin{equation}B(\pi\lambda,I_{i}\rightarrow I_{f})=\int dE_x \ {dB\left(
  \pi \lambda, E_{x}\right)\over dE_x}.
  \end{equation} 
The differential cross sections can also be expressed in terms of equivalent photons, namely,
\begin{equation}
\frac{d\sigma_C (E_{x})}{d\Omega }=\frac{1}{E_{x}}\sum\limits_{\pi \lambda }%
\frac{dn_{\pi \lambda }}{d\Omega }(E_{x},\theta )\sigma _{\gamma }^{\pi
\lambda }(E_{x}),  \label{(1.6)}
\end{equation}
with $\Omega$ being the solid scattering angle. This  is the same as
Eq. (\ref{cross_2}), but rewritten in a simpler form, helping us to immediately see the connection with the cross sections induced
by real photons. 

Figure \ref{vpn} shows a calculation ($E_\gamma \equiv E_x$) for
virtual photons with $E1$ multipolarity,  and three  typical bombarding energies, ``as viewed'' by a lead 
projectile incident on a lead target at impact parameters larger than
$b=12.3$ fm. When the projectile energy  
increases, more virtual photons with larger energies become available. The energy of states probed  are also increased, making it possible to study  giant resonances, lepton and meson production, and the production of heavy particles \cite{BERTULANI1988299}.

\begin{figure}[t]
\begin{center}
{\includegraphics[width=10cm]{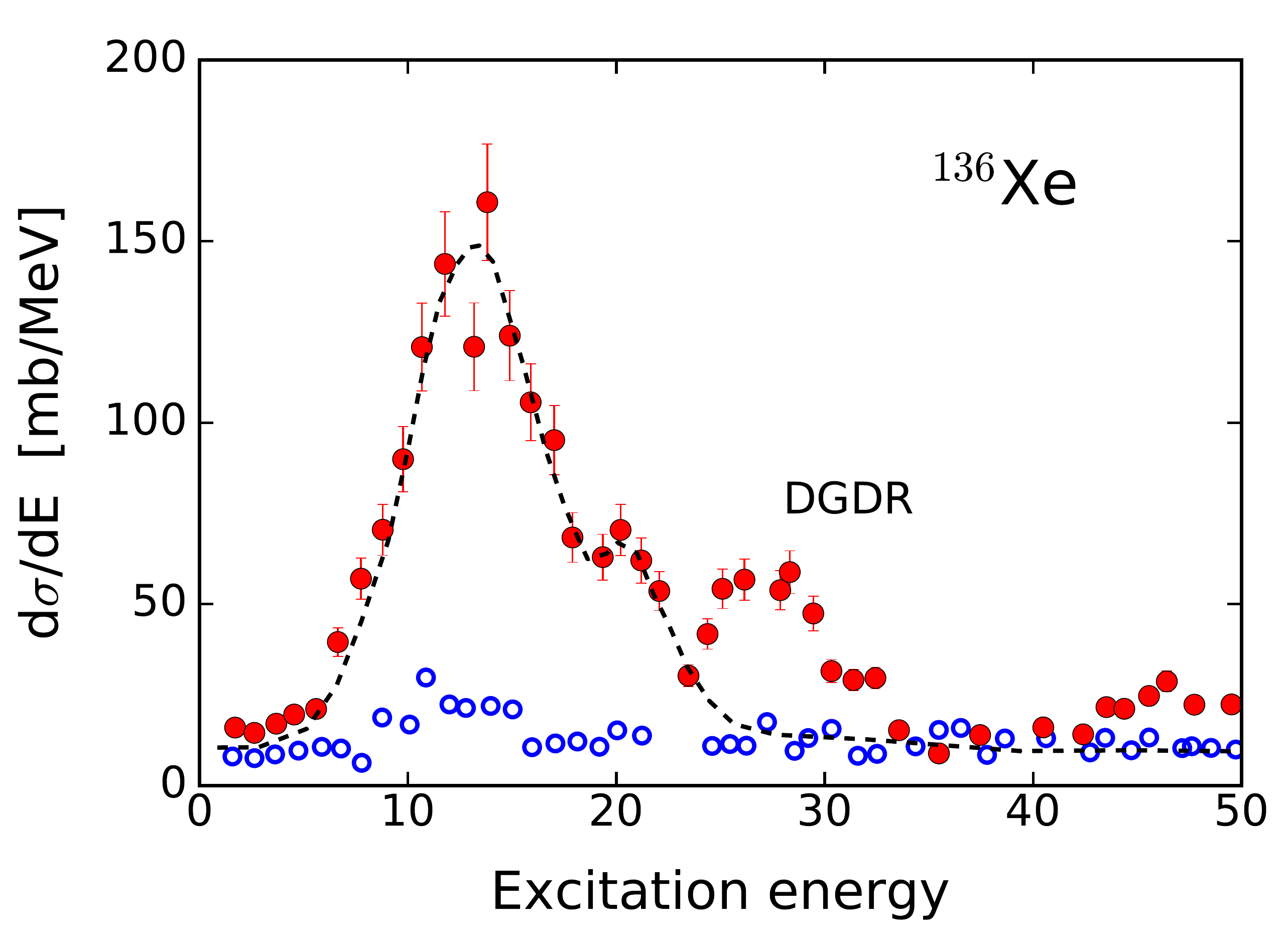}}
\end{center}
\caption{\label{dgdr}
Coulomb excitation cross section for  $^{136}$Xe (700 MeV/nucleon) nuclei impinging on lead (solid circles) and carbon  nuclear targets (open circles). The dashed curve includes  the excitation of isoscalar and isovector giant quadrupole resonances and the isovector giant dipole resonance (IVGDR).  The double giant dipole resonance (DGDR) was clearly identified as a bump in the spectrum at about twice the energy of the IVGDR \cite{SchmidtPRL.70.1767}.}
\end{figure}

A photonuclear
reaction  can access
information complementary to Coulomb excitation. 
For photon energies larger than nucleon separation energies, the photo-absorption cross section displays
characteristic single-particle resonances. For energies in the range of 15-25 MeV, a wide and large peak is observed, known as
the {\it   giant electric dipole resonance} ($E1$ excitation).  Figure
\ref{dgdr} exhibits the  photoabsorption cross section of $^
{136}$Xe at photon energies around the electric dipole giant
resonance and the double giant dipole resonance \cite{SchmidtPRL.70.1767}.

{\it Giant resonances} occur in basically all nuclei along the  periodic table. Their centroid energy
decreases following the phenomenological formula $E_{GDR} \simeq 80/A^{1/3}$
for $A>20 $.  Their widths are almost all in the range between 3.5
MeV and 5 MeV, with few cases reaching up to 7~MeV. They are 
collective excitations with many nucleons participating 
at once. The oscillating electric field of a photon 
effectively induces collective oscillations of protons against
neutrons.
Among the giant resonances, the {\it giant  electric dipole (GDR)  resonance} absorbs one unit of angular momentum ($\Delta l=1 $), e.g., if the nucleus is even-even it is taken to
an $1 ^ - $ state. It is also an  {\it isovector
resonance} because isospin is also changed by one unit
($\Delta T=1 $). Protons and neutrons
vibrating in phase yield {\it isoscalar resonances} ($\Delta T=0 $) and
if in opposite phases they yield  {\it isovector
resonances} ($\Delta T=1 $). The photon excites less effectively giant isoscalar resonances, with $\Delta
T=0 $. Isoscalar  monopole
($\Delta l=0 $) resonances are mostly excited in reactions involving the nuclear interaction. In a {\it giant electric quadrupole resonance} the nucleus
vibrates in an ellipsoidal mode.  In a giant monopole resonance  the nucleus contracts and
expands radially, known as a {\it breathing mode}, also occurring 
in isoscalar and isovector  forms. Monopole resonances are a good probe of
the {\it compressibility of  nuclear matter}.

{\it Magnetic giant resonances} involve {\it
spin vibrations} where nucleons with spin up oscillate out of phase
with nucleons with spin down, also including isoscalar and isovector modes. Charge-exchange reactions are a good probe 
of magnetic resonances induced when a
projectile charge changes down to $Z-1$ or up to $Z+1$, as for eaxmple induced in (p,n), (d,p) and (d,n) reactions. Giant spin-flip resonances are also known as {\it giant Gamow-Teller resonances}.

Giant resonances have also been observed in excited nuclei, first predicted by using the Brink-Axel hypothesis \cite{brink:1955} \cite{axel:1962:PREV}. 
Two giant resonances can be excited simultaneously, e.g., the double giant dipole resonance (or {\it multiphonon
giant resonance}) has been observed  in double charge exchange reactions with pion probes in  $^ {32}$S
\cite{MordechaiPRL.61.531}. Coulomb excitation is perhaps the best probe to excite giant multiphonon resonances,
as predicted in Ref. \cite{BERTULANI1988299} and observed in Ref. \cite{SchmidtPRL.70.1767}. Figure \ref{dgdr} shows a beautiful example of the excitation of the double giant resonance.

\section{Charge exchange reactions}

{\it Charge exchange reactions} are used as a probe to extract the magnitude of {\it Gamow-Teller}, $B(GT)$,  and {\it Fermi}, $B(F)$,  {\it matrix elements}, not accessible in $\beta$-decay experiments \cite{TADDEUCCI1987125}. This technique  relies of the similarity of the interaction in charge-exchange reactions and the spin-isospin operators involved in weak-decay. Using the DWBA approximation, one can show that  the cross section for charge-exchange at small momentum transfers $q$ is proportional to $B(GT)$ and  $B(F)$ \cite{Bertulani:NPA1993:554:493}, 
\begin{equation}
{d\sigma\over d\Omega}(\theta=0^\circ)=\left( \mu \over 2\pi \hbar\right)^2 {k_f \over k_i} N_D|J_{\sigma\tau}|^2 \left[ B(GT) + C_F B(F) \right], \label{tadeucci}
\end{equation}
where  $\mu$ is the reduced mass, $k_i(k_f)$  is the initial (final) relative momentum,  $N_D$ is a correction factor
(accounting for initial and final state interactions), $J_{\sigma\tau}$ is the volume integral of the GT part of the effective nucleon-nucleon interaction, the constant ${C_F} = \left| J_\tau/J_{\sigma\tau}\right|^2$ accounts for possible Fermi excitations, and  $B(\alpha=GT, F)$ is the reduced transition probability for spin-flip and non-spin-flip transitions ($\tau_k$ is the isospin operator). For non-spin flip transitions, it is given by 
$$B(F)= {1\over 2J_i+1}| \langle f ||\sum_k  \tau_k^{(\pm)} || i \rangle |^2,$$ 
and for spin-flip  ($\sigma_k$ is the spin operator),
$$B(GT)= {1\over 2J_i+1}| \langle f ||\sum_k \sigma_k \tau_k^{(\pm)} || i \rangle |^2.$$ 
Small momentum transfers, $q\sim 0$, occur at very small scattering angles, when  $\theta  \ll 1/kR$, where $R$ is a rough measure of the nuclear radius and $k$ is the projectile wavenumber. 

\begin{figure}[t]
\begin{center}
\includegraphics[
width=3.5in]{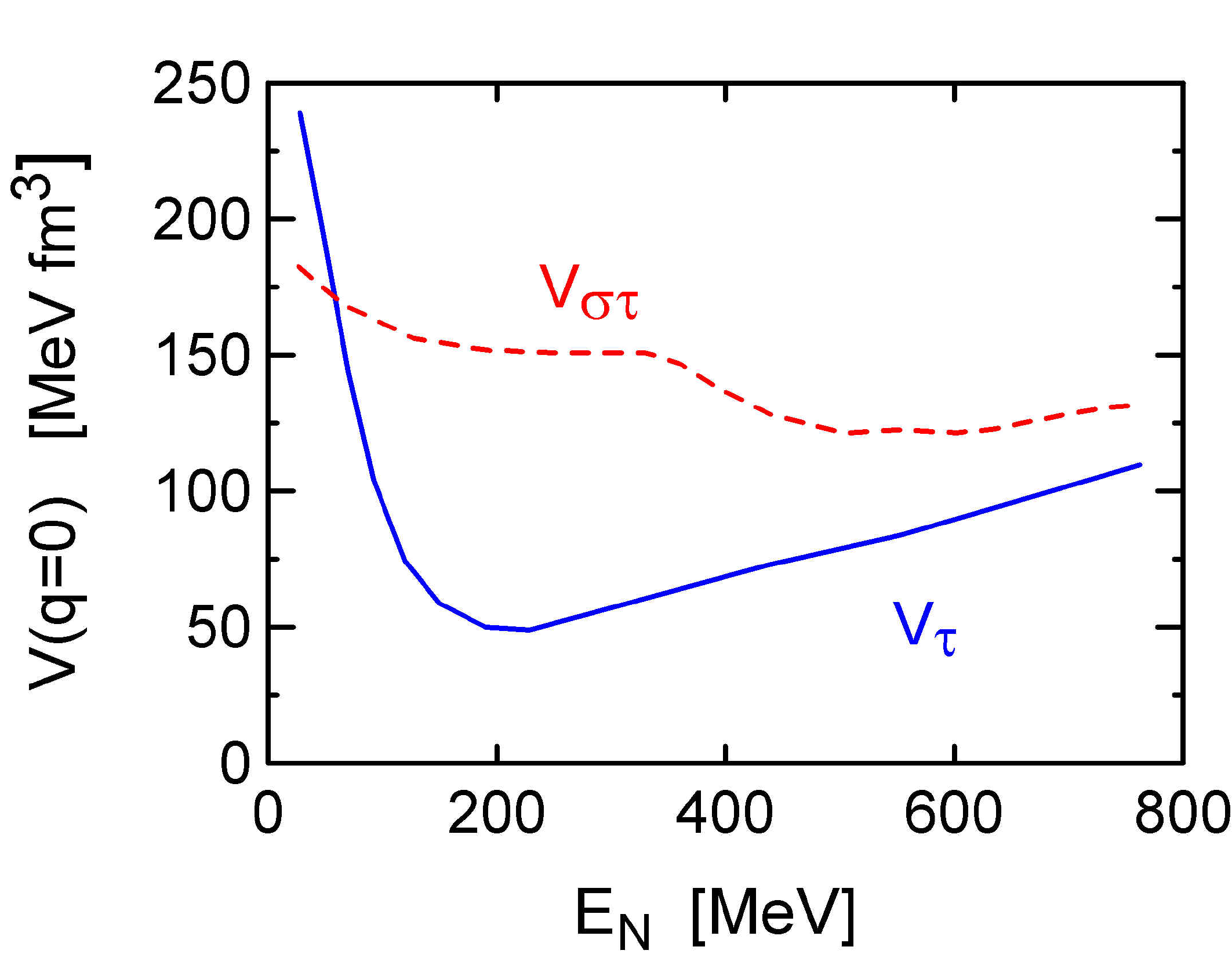}
\end{center}
\caption{Strength of the nucleon-nucleon potential at forward scattering angles. Separate contributions are shown for  the spin-isospin, $\sigma \tau$, and the isospin, $\tau$, components of the interaction as a function of the laboratory energy.}
\label{vnnf}
\end{figure}

Charge-exchange reactions at high energies are due to the exchange of charged pions and rho mesons carrying spin and isospin quantum numbers. Fig. \ref{vnnf} displays the energy dependence of the nucleon-nucleon potential at forward angles. The separate contributions of spin-isospin, $\sigma \tau$, and isospin, $\tau$, parts of the interaction are also shown. One observes that intermediate energy collisions, i.e., around $E\sim 100-300$ MeV, have larger $\sigma \tau$ than $\tau$ contributions. For this reason, this energy range is better suited for studies of the Gamow-Teller matrix elements needed for astrophysics. Therefore, at intermediate energy collisions, 
\begin{equation}
{d\sigma\over dq}(q=0)\sim KN_D|J_{\sigma\tau}|^2 B(GT) , \label{tadeucci2}
\end{equation}
with $K$ being a kinematical constant.  
In the plane-wave Born-approximation the  charge-exchange scattering matrix elements are given by \cite{Bertulani:NPA1993:554:493}
\begin{equation}
{\mathcal M}_{exch}({\bf q})=\left<\Psi_A^{(f)} ({\bf r}_A)\Psi^{(f)}_B({\bf    r}_B) \left| e^{-i{\bf q}\cdot {\bf r}_A }  V_{exch}({\bf q})e^{i{\bf q}\cdot {\bf r}_B} \right|  \Psi_A^{(i)} ({\bf r}_A)\Psi^{(i)}_B({\bf r}_B) \right>, \label{Mexch}
\end{equation}
with ${\bf q}$ being the momentum transfer, $\Psi_{A,B}^{(i,f)}$ the intrinsic wavefunctions of nuclei $A$ and $B$ for the initial $i$ and final $f$ states. ${\bf
  r}_{A,B}$ are intrinsic coordinates of the participating  nucleons  and $V_{exch}$ is charge-exchange part of the nucleon-nucleon interaction  containing spin and isospin operators.  At forward angles and low-momentum transfers,  
${\bf q}\sim 0$, the matrix element (\ref{Mexch}) becomes
\begin{equation}
{\mathcal M}_{exch}({\bf q} \sim 0) \sim V_{exch}^{(0)} ({\bf q}\sim 0) \, {\mathcal M}_A(F,GT)\, {\mathcal M}_B(F,GT)
\, ,
\label{q1}
\end{equation}
where $V_{exch}^{(0)} $ is the volume part of the interaction, and 
$${\mathcal M}_{exch}(F,GT)= \left<\Psi_{A,B}^{(f)}\vert\vert (1, \ \sigma ) \tau \vert\vert   \Psi_{A,B}^{(i)}\right>$$  
are Fermi (F) or Gamow-Teller (GT) matrix  elements for the nuclear transition. One certainly expects deviations from the PWBA and $B(GT)$ values extracted using Eq. (\ref{tadeucci2}) can become inaccurate. This equation has been widely used in the literature, although it is known to fail in some situations. It can be used with caution to infer electron capture, beta-decay, or neutrino scattering response functions in nuclei from charge-exchange reactions \cite{BERTULANI1997237}.  

The validity  of one-step processes in Eq. (\ref{tadeucci})  is a reasonable assumption for $(p,n)$ reactions. But, in  heavy-ion charge-exchange reactions this might not be as appropriate, as shown in Refs. \cite{LenskePRL62.1457,Bertulani:NPA1993:554:493}.  Multi-step processes including the physical
exchange of a proton and a neutron were considered in Ref. \cite{LenskePRL62.1457} and shown to be relevant up to 100 MeV/nucleon. Deviations from Eq. (\ref{tadeucci})  are common under many circumstances  \cite{BERTULANI1997237}. For GT transitions comprising a small fraction of the sum rule, a direct proportionality between $\sigma($p,\ n$)$ and $B(GT)$ values does not exist. Discrepancies have also been observed \cite{WatsonPRL55.1369} in reactions with odd-A nuclei including $^{13}$C, $^{15}$N, $^{35}$Cl, and $^{39}$K and in charge exchange with heavy ions \cite{SteinerPRL76.26}. 

\subsubsection{Double-charge-exchange and double-beta-decay}

\begin{figure}[t]
\begin{center}
\includegraphics[
width=3.5in]{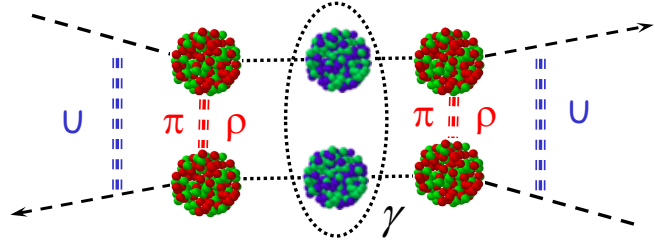}
\end{center}
\caption{Schematic description of a double-charge exchange reaction, where a two-step process is induced by the nucleon-nucleon interaction. The optical potential $U$ is responsible for the elastic scattering of the incoming and outgoing nuclei, whereas the charge exchange is viewed as pion+rho exchange in nucleon-nucleon interactions.}
\label{dcx}
\end{figure}

A more ambitious reaction probe involves {\it double-charge exchange reactions}, seen schematically in Figure \ref{dcx}. They may be used to extract matrix elements for {\it double beta decay in nuclei} for  a number of energetically allowed decays. In the DWBA, amplitude for this process involves the matrix element
\begin{equation}
{\cal M}({\bf k},{\bf k}') = \sum_{\gamma,{\bf k''}} C_{\gamma}\left<\chi_{\bf k'}^{(-)}\left| V_{exch}{1 \over E_{\bf k}-\epsilon_{\gamma,{\bf k''}}-T-V_{exch}}V_{exch}  \right| \chi_{\bf k}^{+}\right> ,
\label{Tdcx}
\end{equation}
where $\chi_{\bf k}$ is the distorted scattering wave in an optical potential $U$, $\bf k$, $\bf k'$ are the initial and final scattering momenta, $\bf k''$ is the momentum of an intermediate state  $\gamma$ with energy $\epsilon_{\gamma,{\bf k''}}$ and $T$ is the kinetic energy operator. $C_{\gamma}$ includes spectroscopic amplitudes of the intermediate states. At forward scattering angles, and using the same approximations as in Eq. (\ref{q1}), a proportionality also emerges between double charge-exchange reactions and double beta-decay processes. The typical cross sections in a single step charge exchange reaction is a few millibarns, whereas a double charge exchange cross section is expected to be less than microbarns \cite{Bertulani:NPA1993:554:493}.

Double beta decay are typically ground state to ground state transitions, accompanied by two neutrino emission,  or by no neutrino emission. In the latter scenario, they place constraints on particle physics beyond the standard model, involving concepts such as lepton number non-conservation, and the neutrino being a Majorana particle, e.g., its own anti-particle. Neutrinoless double beta decay studies involve the neutrino mass and complicated nuclear transition matrix elements. Double beta decays into two neutrinos have been observed \cite{ElliottPRL59.2020}, but  neutrinoless double beta decay still remains elusive in experimental nuclear physics.

Fermi type operators do not contribute appreciably to double beta-decay when neutrinos are emitted,  because the ground state of the final nucleus is not a double isobaric analog of its initial state. Hence, the relevant transitions are of double Gamow-Teller type. In neutrinoless beta-decay Gamow-Teller are expected to be larger than Fermi transitions  \cite{ZHENG1990343}.
Using double charge-exchange reactions as a  probe of double-beta decay matrix elements is now being pursued by many experimental groups (see, e.g., \cite{Matsubara2013,Kisamori:PRL.116.052501,Cappuzzello2018}) not only for assessing neutrinoless double beta-decay but also to study exotic nuclear structures (see, e.g., \cite{LenskePRC98.044620,ShiumizuPRL120.142502}).

\begin{figure}[t]
\begin{center}
{\includegraphics[width=10cm]{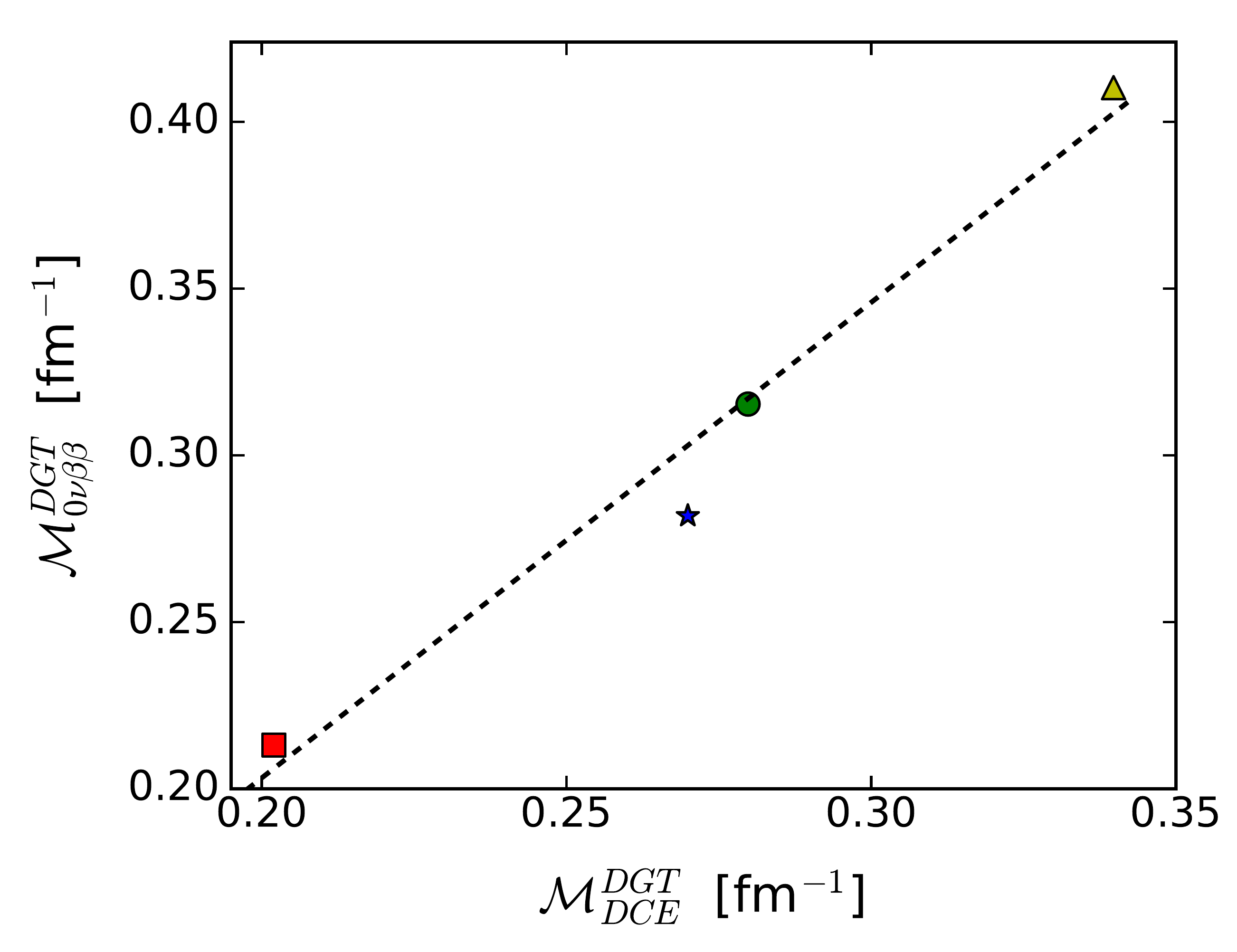}}
\end{center}
\caption{\label{dchexch}
 Correlation between the calculated double charge-exchange (DCE) nuclear matrix elements (NME) for Gamow-Teller (GT) transitions 
and for neutrinoless double beta-decay (0$\nu\beta\beta$) \cite{SantopintoPRC98.061601}. The theoretical results are  for $^{116}$Cd $\rightarrow$ $^{116}$Sn, $^{128}$Te $\rightarrow$ $^{128}$Xe, $^{82}$Se $\rightarrow$ $^{82}$Kr, and $^{76}$Ge $\rightarrow$ $^{76}$Se, respectively.  
}
\end{figure}

Fig. \ref{dchexch} shows that a correlation seems to exist between calculated {\it double charge-exchange} (DCE) nuclear matrix elements (NME) for Gamow-Teller (GT) transitions  and neutrino less double beta-decay (0$\nu\beta\beta$) (adapted from Ref. \cite{SantopintoPRC98.061601}). The calculations have been performed  for $^{116}$Cd $\rightarrow$ $^{116}$Sn, $^{128}$Te $\rightarrow$ $^{128}$Xe, $^{82}$Se $\rightarrow$ $^{82}$Kr, and $^{76}$Ge $\rightarrow$ $^{76}$Se.  The nearly linear correlation can be explained with a transparent reaction theory and if it remains robust, it will open the possibility of extracting neutrinoless double beta-decay NMEs from experimental data on DCE at forward angles. A recent review provides many clarifications for charge-exchange reactions as a probe of nuclear $\beta$-decay \cite{LENSKE2019103716}.

\section{Transfer reactions}

In the scattering of two nuclei, {\it nucleon transfer reactions} occur when a nucleon or a cluster of nucleons initially in a bound state of one of the two nuclei ends up in a bound state of the other nucleus. From the point of view of quantum mechanics this happens because when the two nuclei are close together there is a finite overlap between the tails of the initial and final wave functions. If this overlap is large the cross section will be large compared to other reaction cross sections. For this reason transfer reactions are very selective, in particular for heavy nuclei, and they can be a very powerful spectroscopic tool. One can get information on the {\it spectroscopic factors}  (SF) and/or {\it asymptotic normalization constants} (ANC) and the angular momentum quantum numbers of the wave-functions. The negative single particle binding energies of valence nucleons  are known   when the nucleus mass is accurately determined experimentally. However for nuclei very close to the drip line it might happen than the mass is not so well determined due to the nucleus very short time-life. 

A transfer reaction can be represented as
\begin{equation}
\centering
A_1(a_1+x)+A_2\to a_1+(A_2+x),\label{t1}\end{equation}
where $x$ is the transferred nucleon or cluster.

A simple classical relationship between the binding energies ($\varepsilon_{n1}$ and $\varepsilon_{n2 }$ ) and the incident energy per nucleon $\frac 1 2 m v^2$ indicates that the optimum matching condition \cite{VONOERTZEN198595,Bonaccorso_1987,BRINK197237,DMBB} is 
\begin{equation}
\centering
|\varepsilon_{n1}-\varepsilon_{n2}|=\frac 1 2 m v^2.\label{t2}\end{equation}

Eq. (\ref{t2}) shows that when $\frac 1 2 m v^2$ is very large or/and one of the binding energies is very small, then the most favored final energy for the nucleon/cluster might be positive. In this case one would talk of {\it transfer to the continuum} which is usually called {\it breakup} and that we will discuss in the following section. Furthermore, it has been shown both experimentally \cite{WINFIELD1988345} as well as theoretically \cite{VONOERTZEN198595,Bonaccorso_1987}
 that the transfer probability
increases from the Coulomb barrier up to a maximum at an incident energy $E_{crit}$, 
given approximately by the condition 
\begin{equation}
  \centering  
  E_{crit}=|Q|-V_{CB}/A_{12},
\end{equation}
where $Q$ is the reaction Q-value, $V_{CB}=Z_1Z_2e^2/R_s$ is the value of the projectile-target Coulomb barrier at the strong absorption radius $R_s$ and  $A_{12}=A_1A_2/(A_1+A_2)$.

At high energies the transfer probability decreases with the angular momentum and spin of the initial and final state.
The spin dependence of the transfer probability is influenced both by the reaction Q-value and by the spin coupling factors between initial and final states.
Classical arguments suggest \cite{BUTTLE1971299,BRINK197237} that spin-flip transitions ($j_1=l_1 \pm  \frac 1 2 \to j_2=l_2 \mp \frac 1 2 $ ) are favored at low incident energies
while the opposite occurs at high energies. The inversion from one regime of spin selectivity
to the other occurs when  $E_{inc}\approx E_{crit}$.
On the other hand, because the angular momentum must be conserved, the difference between the initial and final angular momenta must be provided by the relative motion angular momentum, such that \cite{BRINK197237}:
\begin{equation}
    \hbar (\lambda_1-\lambda_2)=\left |\frac {RQ_{eff}} v \right|
\end{equation}
where $Q_{eff}$ takes into account the Coulomb barriers and $R$ is close to the sum of the radii of projectile and target.
If the {\it matching conditions} are not satisfied, the cross sections for transfer are going to be depleted with respect to other competing channels.

\begin{figure}[t]
\begin{center}
       \includegraphics[width=3.3in]{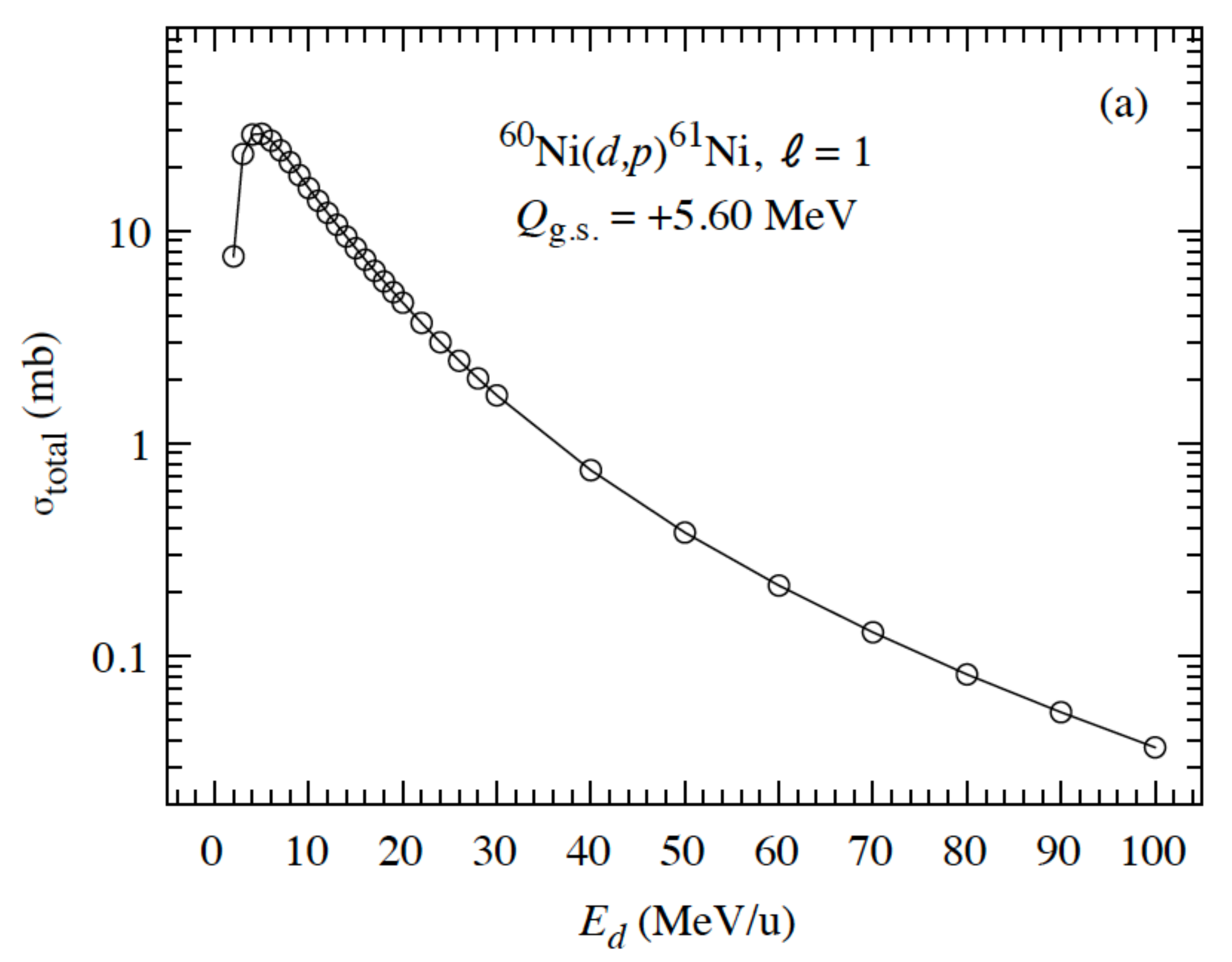}
      \includegraphics[width=3.3in]{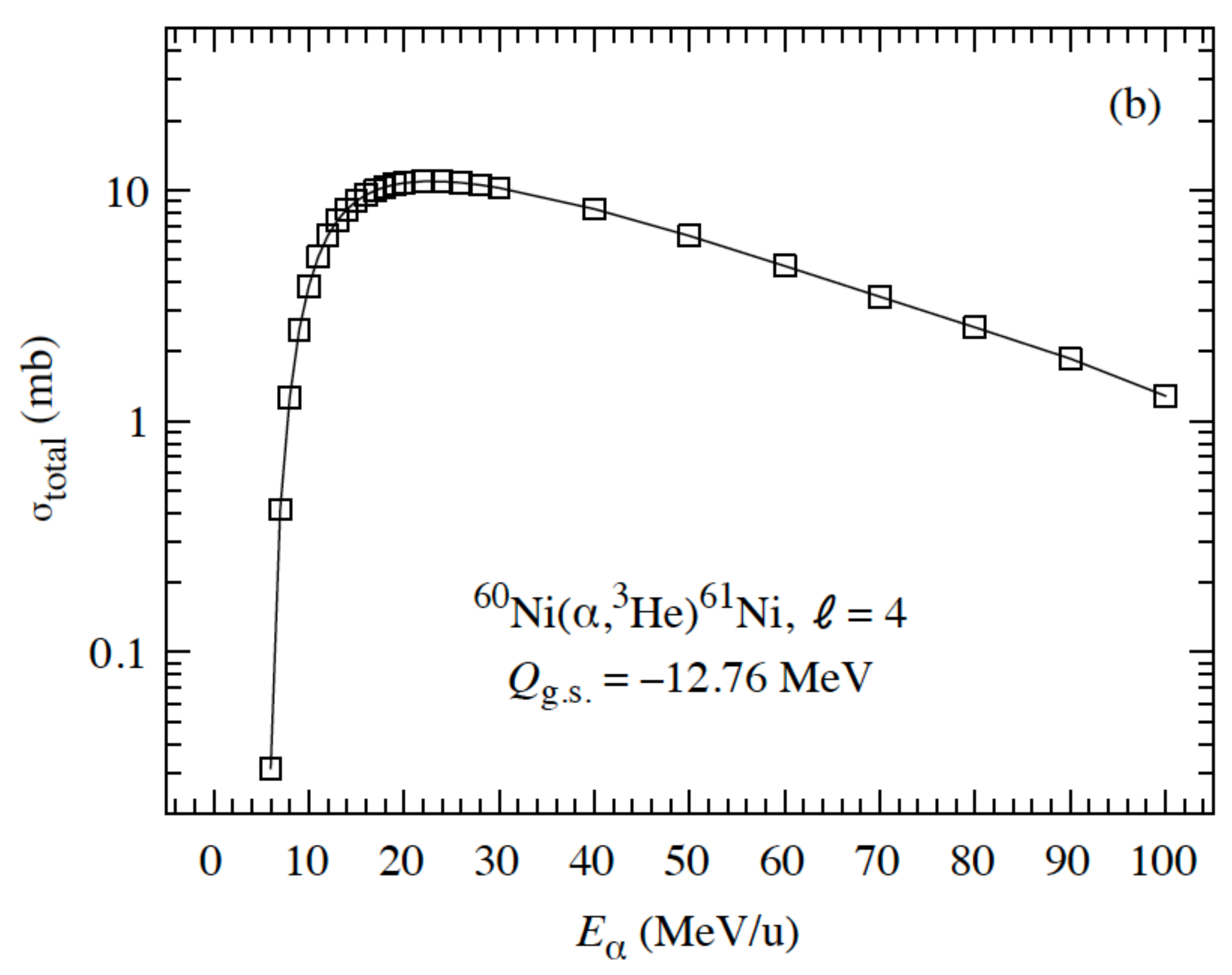}
    \caption{Energy dependence of two reactions showing a peak at the critical energy. See text for details.}
    \label{fig:fig21}
    \end{center}
\end{figure}

\begin{figure}[t]
\begin{center}
   \includegraphics[width=3.7in]{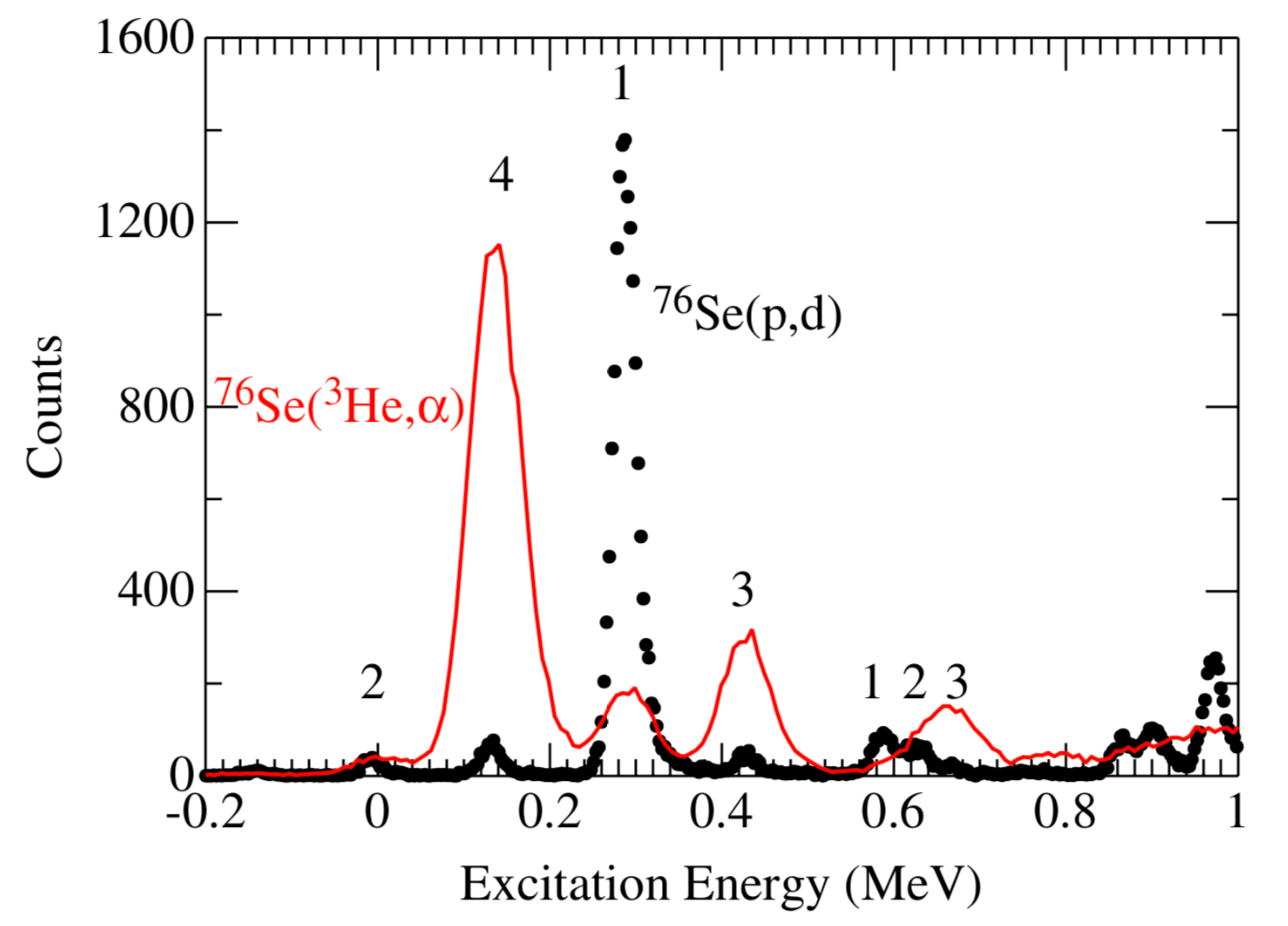}
      \includegraphics[width=3.8in]{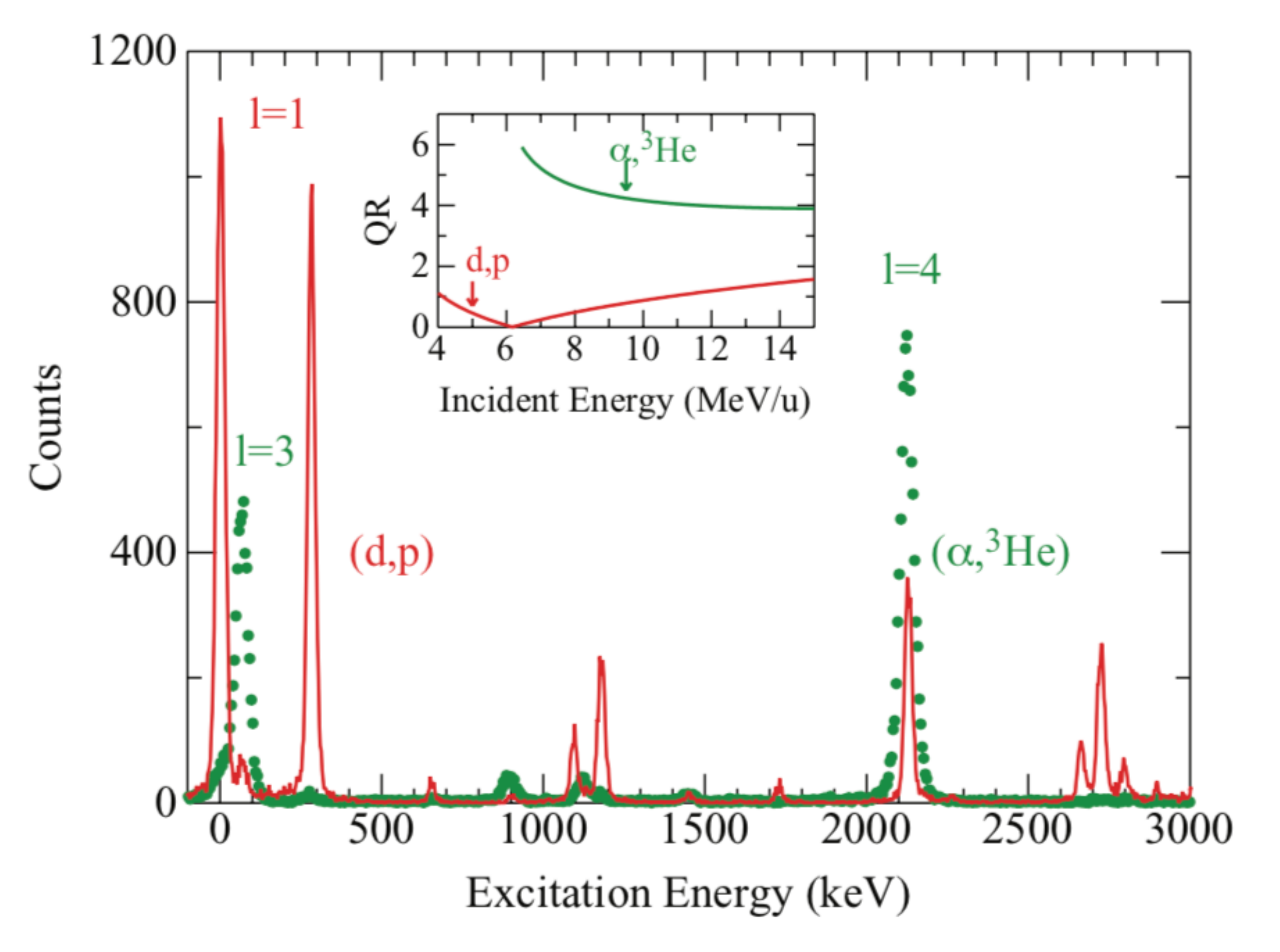}
    \caption{Spectra from the neutron adding (p,d) and ($^3$He,$\alpha$) reactions, top figure and neutron-removing,  (d, p)   and ($\alpha$,$^3$He) reactions, bottom figure. 
Shown in the inset, is the momentum matching  which indicates $R Q_{eff}  \approx 1$ for the (d, p) reaction and $R Q_{eff}  \approx 4 $ for the ($\alpha$,$^3$He) reaction. The ratios of the cross sections between the  two reactions for $\ell=1$ or 4 differ by about two orders of magnitude depending on their $\ell$ value demonstrating the effect of the momentum matching.}
    \label{fig:fig23}
    \end{center}
\end{figure}

Following the discussion of transfer reactions contained in \cite{EMMI},
Figs. \ref{fig:fig21} and  \ref{fig:fig23} show two examples of the momentum matching condition. Figure \ref{fig:fig21} contrasts  the neutron adding (d, p) and ($\alpha$,$^3$He) reactions on $^{60}$Ni, 
while Figure \ref{fig:fig23} compares the neutron-removing (p,d) and ($^3$He,$\alpha$) reactions on $^{76}$Se \cite{Schiffer08} and \cite{Schiffer13}. In the first example, the beam 
energies were about 5 MeV/nucleon for the (d,p) reaction and 9.5 MeV/nucleon for the ($\alpha$,$^3$He) reaction,  both being near the peak cross 
sections, as seen in Fig. \ref{fig:fig21}. These results are consistent with some previous works on heavy ion scattering    \cite{WINFIELD1988345,VONOERTZEN198595,Bonaccorso_1987}.

From Fig. \ref{fig:fig23}, bottom part, the energy spectra for the two reactions, ($d$,$p$) and ($\alpha$,$^3$He), appear  to be very different. The momentum matching condition, shown in the inset, indicates 
$R Q_{eff}  \approx 1$ for the $(d,p)$ reaction and $\approx 4$ for the $(\alpha,^3$He) reaction. The ratios of the cross sections between the two reactions for $\ell$=1 or 4 differ by nearly two orders of magnitude depending on their $\ell$ value. This is an astonishing demonstration that momentum matching occurs.
We thus understand that the cross sections from the respective reactions are low (high) because of poor (good) momentum matching. When the matching is poor, the contributions of more complicated, indirect (multi-step) pathways can contribute more significantly, and the interpretation of the cross section being a simple one-step process becomes  questionable.
As a result the  structure information extracted, such as the spectroscopic factors, for example, will not be reliable.
 
 Transfer observables can be calculated with the DWBA theory or with semiclassical methods which are obtained from the previous when certain conditions apply.
First we derive the formulae for the angular distributions in DWBA theory. The scattering amplitude for a reaction $A_1(A_2,a_1)a_2$ in the center-of-mass frame,  is
\begin{equation}
\label{integration}
T_{\rm DWBA}(\theta,\phi)= -\frac{\mu_\beta}{2\pi\hbar^2} \int\int\chi^{(-)}_{\beta}({\rm\bf k_{\beta}},{\rm\bf r_{\beta}})^{\ast} \left< {\rm a_1},{\rm a_2}\left|V_{eff}\right|{\rm A_1},{\rm A_2}\right> \chi^{(+)}_{\alpha}({\rm\bf k_{\alpha}},{\rm\bf r_{\alpha}}){\rm d\bf r_{\beta}},
\end{equation}
where the functions $\chi_{\alpha}$ and
$\chi_{\beta}$ are distorted waves describing the elastic scattering of the particles in  
the entrance ($\alpha={\rm A_1}+{\rm A_2}$) and exit ($\beta={\rm a_1}+{\rm a_2}$) channels, with momentum and  relative coordinates  ${\rm\bf k_{\alpha,\beta}}$ and ${\rm\bf r_{\alpha,\beta}}$,
respectively. $V_{eff}$ denotes the interaction inducing the transition, and $\mu_\beta$ is the reduced mass in the exit channel.

It follows that the cross section for a {\it single particle (s.p.) state}, with a certain angular momentum and orbital angular-momentum transfer, is
\begin{equation}
\frac{d\sigma(\theta)}{d\Omega}=\frac{v_{\beta}}{v_{\alpha}}
\left|T_{\rm DWBA}(\theta)\right|^2,
\end{equation}
where $v_{\alpha}$ and $v_{\beta}$ are the center-of-mass velocities in the incoming  and outgoing  channels.

In Eq. (\ref{integration}), the  matrix element $\left< {a_1,a_2}\left|V_{eff}\right | {A_1,A_2}\right>$ includes an integral over the internal coordinates of the many-body wavefunctions of the incident and outgoing particles. It is common to assume that $V_{eff}$ does not depend on the internal coordinates.  For example, if $a_2$ emerges from the addition of one neutron to the target $A_2$, one needs to perform the {\it overlap integral}
\begin{equation}
\label{eq:ABoverlap}
 \int d\xi \; \Psi^{*}_{a_2} (\xi, {\bf r}) \Psi_{A_2}(\xi)    \equiv \psi^{\ell,j}_{A_2a_2}({\bf r}) ,
\end{equation}
where $\xi$ stands for the internal coordinates of $A$ and ${\bf r}$ that of the additional neutron. This overlap integral 
is proportional to the probability amplitude to find the state $a_1$ when a nucleon is removed from $A_1$. Generally, $\psi^{\ell,j}_{A_1a_1}({\bf r})$ is not normalized to one. Its normalization yields the so-called {\it spectroscopic factor},
\begin{equation}
\label{eq:norm_overlap}
\int d {\bf r} |\psi^{\ell,j}_{A_1a_1}({\bf r})|^2 = S^{\ell,j}_{A_1a_1}.
\end{equation}

In practical calculations using the DWBA, the overlap function is often approximated by a {\it single-particle ($s.p.$) wavefunction}, obtained from the solution of a Schr\"odinger equation with a mean-field potential (such as a Woods-Saxon type), with the appropriate separation energy and quantum numbers  $\ell,j$.  Since the $s.p.$ wavefunction is unit normalized, one writes
\begin{equation}
\label{eq:sp_overlap}
\psi^{\ell,j}_{A_1a_1}({\bf r}) \approx \sqrt{S^{\ell,j}_{A_1a_1}} \psi^{\ell,j}_{sp}({\bf r}),
\end{equation}
where $\psi^{\ell,j}_{sp}({\bf r})$ is the $s.p.$ wavefunction.

With explicit account of the angular momentum, additional {\it Clebsh-Gordan coefficients} appear in the formalism. In addition, if  the isospin formalism is used to express the states $A$ and $B$,  another {\it isospin coefficient} ($C$) appears. They are sometimes  singled out from the definition of the spectroscopic factors and hence written explicitly as ($C^2 S$). The use of Clebsch-Gordan coefficients in the context of transfer reactions in thoroughly discussed in Ref.~\cite{Schiffer69} and an explicit example is given in Ref.~\cite{Szwec16}. $C^2$ is often taken as one and not discussed at length, or maybe intentionally or unintentionally ignored.

If the $s.p.$ overlap Eq.~(\ref{eq:sp_overlap}) is included in the scattering amplitude, Eq.~(\ref{integration}), one may write the differential cross section as  a $s.p.$ cross section multiplied by a respective spectroscopic factor, i.e.,
\begin{equation}\label{abss}
\frac{d\sigma(\theta)}{d\Omega} = \omega \, (C^2 S_i) \left . \frac{d\sigma(\theta)}{d\Omega} \right |_{\rm sp},
\end{equation}
where $S_i$  is the spectroscopic factor for the specific state $i$.  Notice that, if both the projectile and target overlap functions are written in terms of $s.p.$ overlaps, a product of the corresponding spectroscopic factors will appear in Eq. (\ref{abss}). An additional {\it statistical factor} $\omega$ is needed, which is $(2j+1)$ for adding, and one for removing nucleons.

The angular distributions obtained in this way are characterized by oscillations that depend on the transferred angular momentum and thus their measure has been  routinely used to determine $l$-values for the final (initial) $s.p.$ states. This is most easily done  in (d,p) and ($\alpha$,$^3$He) reactions because in this case the initial angular momentum is zero. On the other hand, heavy ion reactions are useful when one wants to study final states of high spins because they are  favored in this case due to the large relative angular momentum. However, in heavy ion reactions there are other channels competing with transfer thus the underlying core-target interaction is quasi-elastic and coupled channels calculations are necessary. For these reasons, at present most of the time one prefers to use (d,p) reactions.

\section{Breakup reactions}

In the previous section we have argued that {\it breakup reactions} have the same physical origin as transfer reactions and that they become dominant when following  Eq. (\ref{t2}) the most favored nucleon or cluster final energy is positive. According to quantum mechanics (QM),  positive energies give rise to a continuum spectrum. Therefore, breakup can be seen as a transfer reaction in which all final energies are possible. In transfer reactions the nucleon final state wave-function is determined only by the final nucleus potential. On the other hand in breakup reactions the particle final state is in the continuum, the particle  will have final state interactions with both the initial and final nucleus potentials, thus in principle its wave-function should reflect both these potentials. Clearly, breakup is more complicated than transfer but at the same time can give more information on the two interacting nuclei and it all depends on the measured observables.

\begin{figure*}[h]
\centering
\includegraphics[scale=.4]{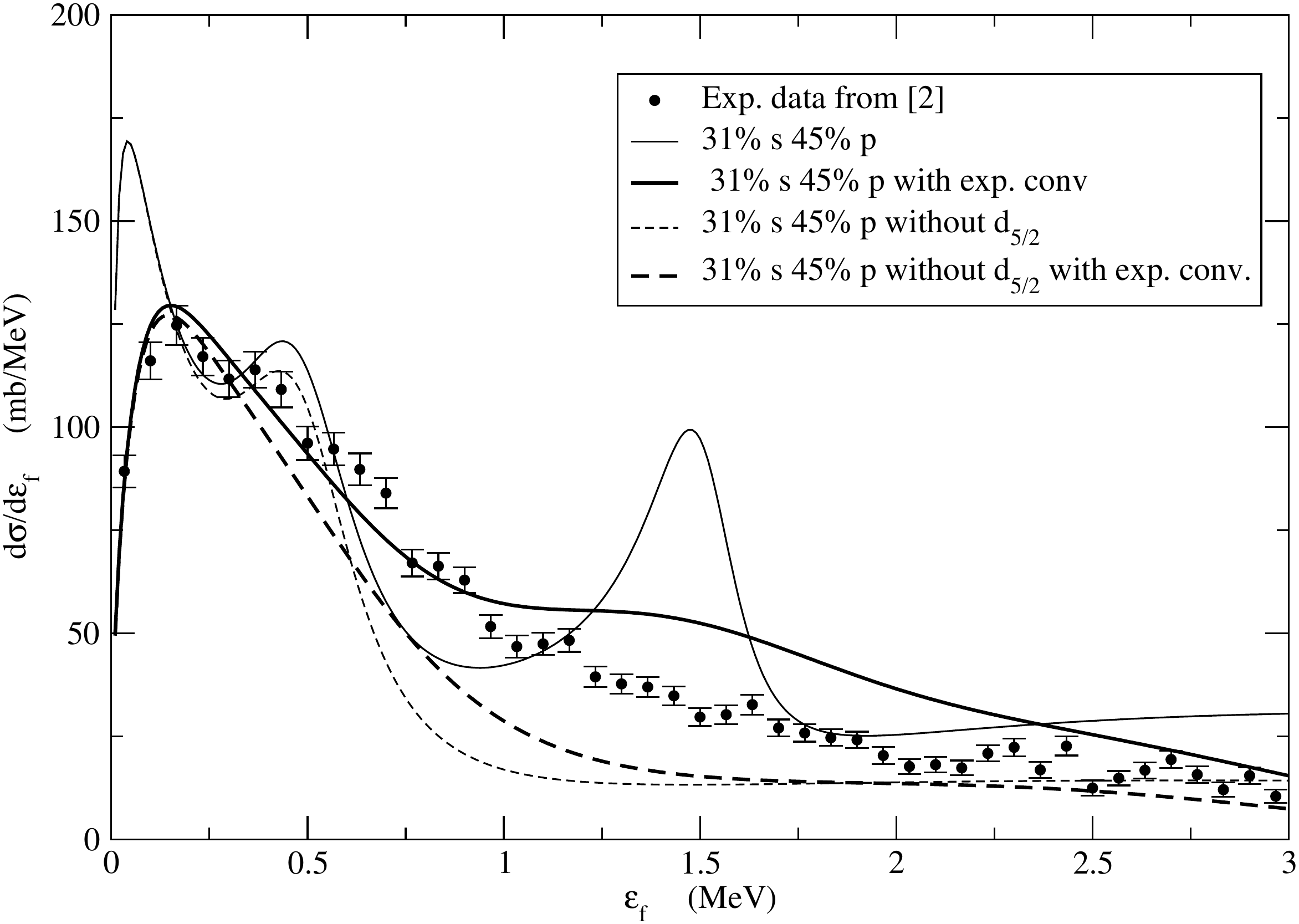}
\caption { Relative-energy spectrum n-$^9$Li following $^{11}$Li projectile fragmentation in the reaction \cite{BLANCHON2007303,SIMON2007267}  $^{11}$Li($^{12}$C,X)$^{9}$Li+n at 264 MeV/nucleon.}
\label{f7}
\end{figure*}

In an  {\it exclusive reaction}, the breakup particle is measured in coincidence with the core of origin, then one has access to the continuum spectrum of the original nucleus and the energy spectrum will provide information on resonance states of the projectile. In such a case, the mechanism is the same as inelastic excitation to the continuum. In Fig. \ref{f7} we give an example for the reaction \cite{BLANCHON2007303,SIMON2007267}  $^{11}$Li($^{12}$C,X)$^9$Li+n at 264 MeV/nucleon. The data and theoretical analysis give information on the low lying resonances of the unbound nucleus $^{10}$Li and on the corresponding components of the two-neutron halo nucleus $^{11}$Li ground state wavefunction.

\begin{figure}[h!]
\vskip 15pt
\centering
\includegraphics[scale=.8]{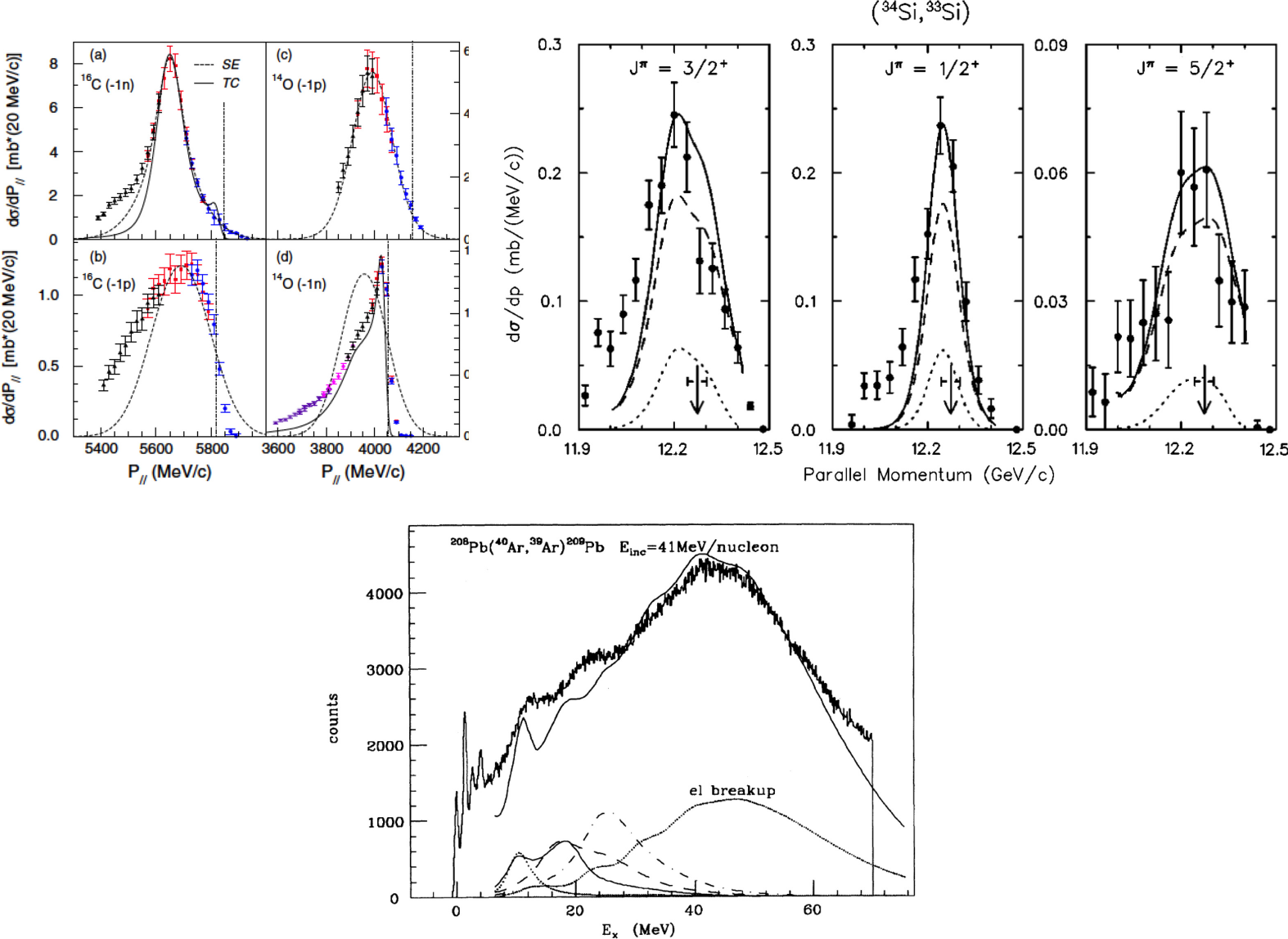}
\caption {Examples of momentum and energy spectra taken from Refs. \cite{isa,end,Flavigny:2012}. }
\label{ffig6}\end{figure}

In an {\it inclusive reaction}, only the original core is measured. A continuum energy spectrum is obtained which contains mainly information on the final state interaction of the breakup particle with the target. If the target final state is not determined, the spectrum contains the effect of both elastic and inelastic scattering similarly to what discussed in previous sections for a beam of free particles. However, because in the initial state the particle is bound the treatment needs special care. Some examples are given in Fig. \ref{ffig6}. On the top part of the figure the spectra are given as a function of the core final momentum. Their shape and width give information on the angular momentum and separation energy of the nucleon in the initial state. The figure at the bottom shows an energy spectrum for the target plus one neutron. It is obtained from the projectile-core  energy ({\it missing mass}) measurement after the collision. One can see the presence of low lying resonances as well as a large bump due to the re-scattering of the nucleon on the target. The peak of the breakup bump or of the simple spectra at the top of the figure corresponds to the nucleon optimal final energy according to Eq. (\ref{t2}).

Breakup can be seen to happen because the nuclear target potential perturbs the initial bound state. When the target is very heavy and the initial state weakly bound there is also another mechanism known as {\it Coulomb breakup} \cite {alder-winther,bertulani:1985:NPA}. It is manly due to the recoil of the core which generates a dipole potential such that there is an effective Coulomb force acting on the breakup particle. Typical measurements consist in  taking in coincidence the breakup particle and its core. However because the  effect of the nuclear potential  is always present the formalism must take care of the interference  between the two mechanisms \cite{me}. For example, Coulomb breakup must be taken into account  in reactions involving the weakly bound deuteron unless the other nucleus involved is very light.

 \begin{figure}[h!]
\centering
   \includegraphics[width=3.5in]{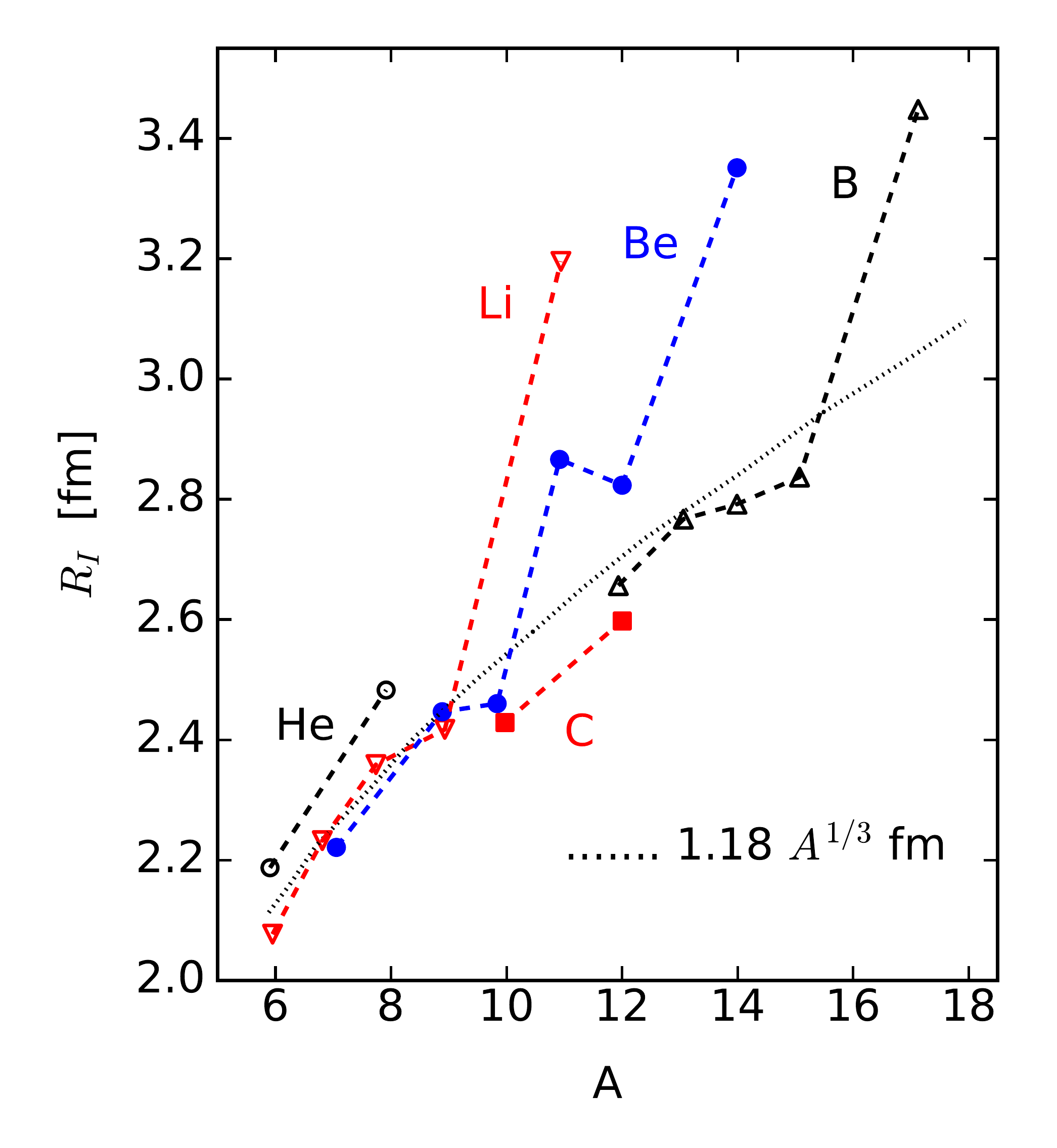}
\caption {Experimentally extracted  interaction radii of light nuclei, determined from the interaction cross sections. The interaction radius (R$_I$) is defined by $\sigma_I=\pi( R_I^P+R_I^T)^2$, where P and
T denote the projectile and target, respectively. A sudden increase of matter radii are seen for nuclei near neutron dripline \cite{Tanihata:96}.}
\label{figR}
\end{figure}
     
Breakup reactions have been extensively studied from the early 80s when it was noticed that in a  systematic of reaction cross sections of Helium, Beryllium and Lithium projectiles in correspondence to some isotopes ($^{6}$Be,$^{11}$Be,$^{11}$Li) the cross sections showed an unexpected enhancement \cite{TANIHATA1985380} (see Fig. \ref{figR}). This is due to the weakly bound valence neutrons and thus to the  strong effect of the breakup channel. For the same nuclei and the same reason, the effect of breakup can also be seen on the elastic  scattering angular distributions, in the form of a depletion of elastic scattering cross sections with respect to neighboring nuclei, as shown by Fig. \ref{figang}.

The theoretical treatment of breakup is in principle very complicated because, at least in DWBA, an expression like Eq. (\ref{integration}) must be solved with a three-body final state which would take into account the nucleon-core, nucleon-target, core-target nuclear potentials plus the core-target Coulomb interaction and, if the breakup particle or cluster is charged, its Coulomb interaction with core and target. However, there is possibly a simplification due to the fact that large breakup  cross sections are measured only for peripheral reactions which apart from the breakup are otherwise quasi-elastic. At small impact parameters instead  channels other than breakup take over and thus the exact value of the breakup cross section is not important.  In such circumstances the core and nucleon behaviors can be decoupled, the so called {\it core-spectator model} applies and the final cross section can be represented semiclassically by an integration over core-target impact parameters
\begin{equation}\frac{d\sigma _{-n}}{d\zeta}=C^2S
   \int d^2{\bf b_c}   |S_{CT}({\bf b_c})|^2 {\frac {d P_{-n}({\bf b_c})} {d\zeta} }.\label{totx}
\end{equation}

In Eq. (\ref{totx}) the variable $\zeta$  can be the nucleon final energy in the continuum, if $\varepsilon_{n2}>0$  in Eq. (\ref{t2}) and/or the nucleon relative momentum with respect to the core or target given
by 4-energy momentum conservation (see for example \cite{firk}) and the relative Jacobian.  The differential cross section $d{\sigma_{-n}/ {d\zeta}}$  then becomes directly comparable to the measured momentum distributions function of P$_{//}$, the core parallel momentum, as in the spectra at the top of Fig. \ref{ffig6}. If a shell model Woods-Saxon wave function is used for the initial  nucleon wave function,  $C^2S$ is the spectroscopic factor of the initial state exactly as in Eq. (\ref{abss}). In the core spectator model the  breakup  cross section is obtained  by integrating the differential breakup probability  on the core-target impact parameter $b_c$ by  weighting it with the probability $|S_{CT}(b_c)|^2$ that the  measured core
has survived  ``intact" the scattering. Breakup reactions substitute transfer as spectroscopic tools whenever in Eq. (\ref{t2}) the most favorite final energy is positive $\varepsilon_{n2}>0$. This typically happens when
${\varepsilon_{n1}}<<\frac 1 2 m v^2$.

In the following we will give three expressions for the breakup probability $d P_{-n}(b_c)/ d\zeta $ which would be used to calculate spectra as those in Figs. \ref{f7} and \ref{ffig6}, respectively.
The energy spectrum of a breakup nucleon with respect of its core of origin is given by an inelastic excitation-like expression,
\begin{eqnarray}
{dP_{in}\over d\varepsilon_{n2}}=\mathcal{C}{1\over 2l_1+1}\Sigma_{m_1,m_2}
 |1-\bar S_{m_1,m_2}|^2 |I_{m_1,m_2}|^2, \label{8}
\end{eqnarray}
where $\mathcal{C}$ contains various kinematical factors, $\bar S_{m_1,m_2}=\exp[2i(\delta+\nu)]$ is a nucleon-core off-the-energy-shell S-matrix which depends on the nucleon-core phase shift  $\delta$ but contains also an extra phase  $\nu$ due to the initial bound state.  $|I|^2\sim {e^{-2\gamma b_c}/ b_c^3}$ 
 can be interpreted as an inelastic-like form
factor and it is interesting to compare it to the transfer to the  
continuum form factor ${e^{-2\eta b_c}/ 
b_c}$ given in the following expression. The inelastic form factor decreases with  
the impact parameter much faster than the transfer/breakup form
factor. This is a  well known characteristic for final bound states  
\cite{bwb} and it would be interesting to see that it persists for final continuum states in future studies.  

The nucleon-target energy spectrum can be calculated  with
\begin{eqnarray} 
\frac{dP_{-n} }{d\varepsilon_{n2}}\approx \frac{1}{2}\Sigma_{j_2}(2j_2+1)(|1- S_{j_2} |^2+1-|S_{j_2} |^2)
\frac{e^{-2\eta b_c}}{2\eta b_c}
\mathcal{F}_{n2,n1}, \label{dpde}
\end{eqnarray} 
where $S_{j_2}$ is the free particle S-matrix obtained with  a  proper nucleon-target optical potential, $\eta$ is a kinematical factor and ${\mathcal{F}}_{n2,n1}$ contains various initial and final states kinematical and spin variables. 
$e^{-2\eta b_c}/{2\eta b_c}$ can be seen as a breakup form factor. The two terms proportional to $|1- S_{j_2}|^2 $ and $(1- |S_{j_2}|^2)$ represent the elastic and inelastic re-scattering of the nucleon on the target. These are indicated by the dotted and dashed curves in the top RHS spectra of Fig. \ref{ffig6}.

If the eikonal approximation is used, the previous equation becomes
\begin{equation}
\frac{dP_{-n}({\bf b_c}) }{d{k_1}}=\int d^2{\bf b}_{n2}(|1-  S({\bf b}_{n2}) |^2+1-|
({\bf b}_{n2})|^2)|\tilde \psi_{n1}(|{\bf b}_{n2}-{\bf b}_c|,k_1)|^2,\label{28}
\end{equation}
where, as in the transfer case, $\psi_{{n1}}$ is the initial state single particle wave function. In the previous expression it enters via its one-dimensional Fourier transform with respect to $k_1 =(\varepsilon_{n2}-\varepsilon_{n1}-{1\over 2}mv^2)/\hbar v$ which  is the nucleon parallel momentum component with respect to the core. Energy and momentum conservation implies that the  core parallel-momentum distribution in the laboratory is
$$
P_{//}= \sqrt{(T_p+\varepsilon_{n1}  -\varepsilon_{n2})^2+2M_r(T_p+\varepsilon_{n1}-\varepsilon_{n2})},
\label{29}
$$
such that a measurement of the latter gives a direct information on the momentum distribution of the valence particle in the initial state of the projectile. $T_p$ is the projectile kinetic energy and $M_r$ the residual mass. Thus, inserting Eq. (\ref{28}) in  Eq. (\ref{totx})  the cross section differential with respect to the intrinsic parallel momentum in the core is obtained in the full eikonal formalism. Finally, by transforming in terms of the core {\it parallel momentum distribution} $P_{//}$, the cross section becomes comparable to the measured spectrum.

 \section{Conclusions}
In this review, we have chosen to discuss only a small subset of properties and features of direct nuclear reactions.   Many subjects were left out and only very general aspects for specific cases have been discussed. Direct nuclear reactions have been a cornerstone tool to probe nuclear structure, and continuously are used in new studies involving radioactive nuclear beams. In fact, nucleon removal and  breakup reactions have been used for a long time  to infer basic properties of nuclei, leading to some surprises, as shown in Figure \ref{figR}, where the large nuclear matter extension in the so-called {\it halo nuclei} was first identified. This finding was the seed of a new era in nuclear physics, allowing for the development of new theoretical formulations for direct nuclear reactions involving weakly bound nuclei.  Direct nuclear reactions is still a very active field in nuclear physics both as an experimental tool of choice and as a playground for new ideas and developments in nuclear reaction theory.

\section{Acknowledgments}
This work has been supported in part by the US DOE grant No. DE-FG02-08ER41533.

\end{document}